%% file: main.tex
\documentclass[reprint,
nofootinbib,
 amsmath,amssymb,
 aps,
floatfix,
showkeys
]{revtex4-2}

\usepackage{footnote}
\usepackage{graphicx}
\usepackage{dcolumn}
\usepackage{bm}

\usepackage[T1]{fontenc}
\usepackage{newtxtext}

\usepackage[letterpaper, left=1in, right=1in, bottom=1in, top=0.75in]{geometry}

\usepackage{multirow}
\usepackage{enumitem}
\usepackage{booktabs}
\usepackage{color, colortbl}
\usepackage[ruled,vlined,linesnumbered]{algorithm2e}
\SetAlFnt{\footnotesize}
\SetAlCapFnt{\footnotesize}
\SetAlCapNameFnt{\footnotesize}
\usepackage{tikz}
\usetikzlibrary{calc, angles, optics, positioning, shadows.blur, arrows, matrix, decorations.markings, math,arrows.meta}

\newcommand{\visibility}{\textsc{Vis\-i\-bil\-i\-ty}\xspace}
\usepackage{graphicx}
\graphicspath{{figures/}}

\usepackage{amsmath}
\usepackage{amssymb}

\usepackage[caption=false,font=footnotesize]{subfig}

\newcommand{\Bigo}[1]{\mathrm{O}\mathord{\left(#1\right)}\xspace}

\newcommand{\mplea}{{\upshape$(\mu + \lambda)$~EA}\xspace}
\newcommand{\mplga}{{\upshape$(\mu + \lambda)$~GA}\xspace}
\newcommand{\xpxea}[2]{{\upshape$(#1+#2)$~EA}\xspace}

\begin{document}

\title{Jones Matrix Characterization of Optical Elements via Evolutionary Algorithms}

\author{Alejandra {De-Luna-Pamanes}}
\email{A01281001@itesm.mx}
\author{Edgar {Covantes Osuna}}
\email{edgar.covantes@tec.mx}
\author{Dorilian {Lopez-Mago}}
\email{dlopezmago@tec.mx}
\affiliation{%
Tecnologico  de  Monterrey,  School  of  Engineering  and  Science,Ave.  Eugenio  Garza  Sada  2501,  Monterrey,  N.L.  64849,  Mexico
\\
}%

\begin{abstract}
\textit{Abstract}---Jones calculus provides a robust and straightforward method to characterize polarized light and polarizing optical systems using two-element vectors (Jones vectors) and $2 \times 2$ matrices (Jones matrices). Jones matrices are used to determine the retardance and diattenuation introduced by an optical element or a sequence of elements. Moreover, they are the tool of choice to study optical geometric phases. However, the current sampling method for characterizing the Jones matrix of an optical element is inefficient, since the search space of the problem is in the realm of the real numbers and so applying a general sampling method is time-consuming. In this study, we present an initial approach for solving the problem of finding the eigenvectors that characterize the Jones matrix of a homogeneous optical element through Evolutionary Algorithms (EAs). We evaluate the analytical performance of an EA with a Polynomial Mutation operator and a Genetic Algorithm (GA) with a Simulated Binary crossover operator and a Polynomial Mutation operator, and compare the results with those obtained through a general sampling method. The results show that both the EA and the GA out-performed a general sampling method of 6,000 measurements, by requiring in average 103 and 188 fitness functions measurements respectively, while having a perfect rate of convergence.
\end{abstract}

\date{\today}


\maketitle

\section{Introduction}
Light is a natural phenomenon that can be detected with the human eye, but more so, it is a complex phenomenon that carries spatial and temporal information. It is a form of energy composed of an electromagnetic field that varies through space and time. One property of light is that of polarization, which describes the temporal variation of its electromagnetic field. Polarization can be manipulated with the help of optical elements so that, as light travels through a given optical element, the electromagnetic field acquires a phase that depends on the properties of the element. This acquired phase can be decomposed into what is called a dynamic phase and a geometric phase. The former relates to the average optical path length, and the geometric phase is related to changes in the state of polarization~\cite{Pancharatnam}.

Polarized light is commonly used to characterize the properties of optical elements. The different methods that analyze the interaction between optical elements and the polarization of light belong to polarimetry. Formally, polarimetry is the science of polarization measurements, so it refers to various methods and techniques utilized to measure and analyze the physical properties related to the polarization of light and its transformations due to the interaction with optical elements~\cite{Gil}. Polarimetry has a wide range of applicability in different fields, going from astronomy to biomedical diagnostics~\cite{polarimetryapps}. Sample measuring optical arrangements are physical arrangements that are used to analyze optical elements by means of a polarimetry technique. A sample measuring optical arrangement consists of a light source, a polarization state generator, the optical elements that will be analyzed, an analyzer and the necessary electronics to analyze the gathered information. 

Optical elements and light's polarization can be described mathematically using either the Jones or the Mueller-Stokes calculus~\cite{Chipman}. The Jones complex-elements vectors representation allows for the inclusion of a global phase, composed of the dynamic and the geometric phase. On the other hand, despite of the well-known advantages of the Mueller-Stokes real-elements vectors representation, it excludes the global phase that is described in the Jones calculus~\cite{Garcia-Caurel,Savenkov}. Another limitation of the Mueller-Stokes approach is that the Mueller matrices composed by 16 real elements structure the information in a complicated manner, since the physical meaning of the elements is not straightforward, and new parameters must be introduced in order to do so. 

This complexity in the interpretation of the Mueller matrices elements creates an absence of a complete understanding of the results which translates into a limitation in the Mueller polarimetry techniques~\cite{LimitationsMuellerPolarimetry}. Furthermore, a great amount of research has been done to deal with the definition of Mueller matrices that represent real-world optical elements, in contrast to Jones matrices for which every matrix can represent a real-world polarization element~\cite{conditions}. Nonetheless, the Jones polarimetry has been greatly ignored, so almost no work has been devoted to the exploration or development of a Jones polarimetry technique.

Recently, Garza-Soto et al.~\cite{Garza-Soto} proposed a traditional search method to characterize the Jones matrix of an optical element. The proposed method takes advantage of the Jones matrix representation, given that a matrix can be reconstructed from its eigenvectors and eigenvalues. Therefore, optical elements can be characterized through their eigenvectors, also known as eigenpolarizations~\cite{Lopez-Mago}. Interestingly, the eigenpolarizations of a Jones matrix can be orthogonal or non-orthogonal. If the eigenpolarizations are mutually orthogonal, the Jones matrix is said to be homogeneous. Otherwise, the Jones matrix is said to be inhomogeneous.

\begin{figure}[htp]
    \centering
    \subfloat[]{
    \centering
        \includegraphics[width=0.46\linewidth]{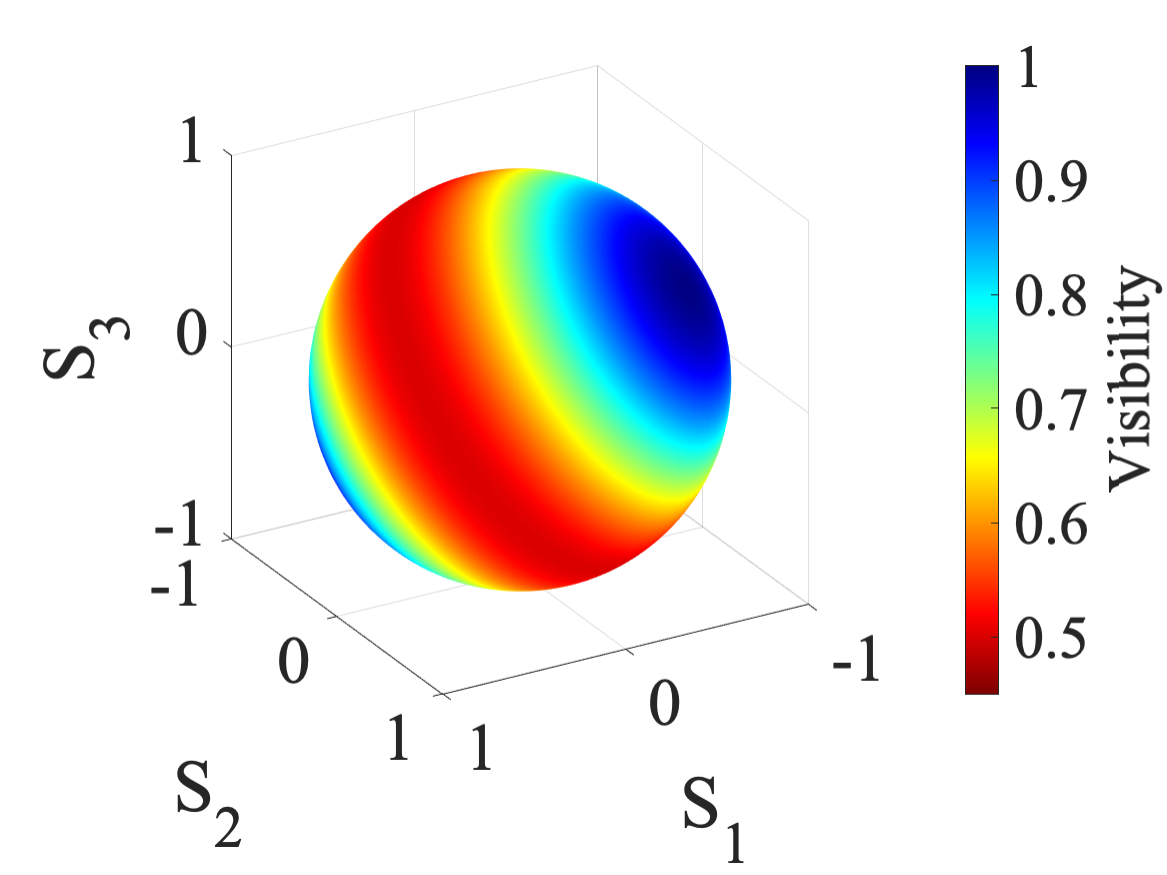}
    \label{fig:homogeneous}
    }
    \subfloat[]{
    \centering
        \includegraphics[width=0.46\linewidth]{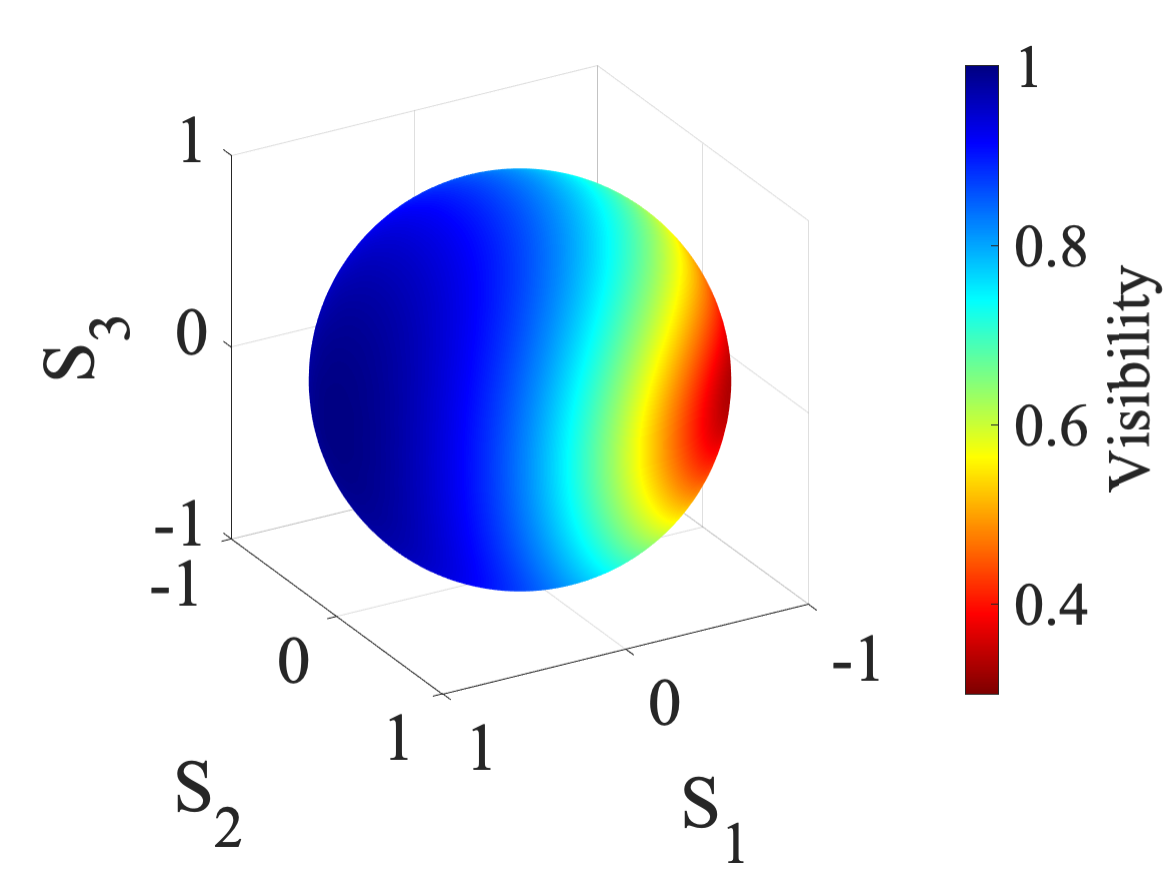}
    \label{fig:inhomogeneous}
    }
    \caption{Numerical simulation of the visibility from the interference pattern between two optical beams, where one of them has interacted with a polarizing optical system. (a) This is the result when considering a homogeneous system. (b) This is the inhomogeneous case.}
    \label{fig:Poincare-HI}
\end{figure}

Figure~\ref{fig:Poincare-HI} shows the interferometric visibility for arbitrary homogeneous and inhomogeneous optical elements over the Poincaré sphere representing the input polarization state. In an interferometric experiment (as shown below in Fig.~\ref{fig:arrangement}), the ``reference beam" interferes with the ``probe beam" that interacts with the sample. The resulting interference contains information about the sample which can be extracted by measuring the visibility of the interference pattern. Figure~\ref{fig:homogeneous} illustrates that homogeneous optical elements have orthogonal maximums instead of the inhomogeneous case in Fig.~\ref{fig:inhomogeneous}, where the maximums' locations are non-orthogonal. Consequently, their location is not so clear at first sight. 
The characterization of inhomogeneous optical elements may require the finding of a pair of vectors with conflicting objective values corresponding to the states of maximum and minimum visibility. Thus, inhomogeneous matrices have more complex properties and are still a subject of study today~\cite{Gutierrez-Vega2020}. On the other hand, the theoretical foundations for homogeneous optical elements is well understood in such way that they can be characterized by finding only one eigenvector of the system corresponding to a maximum value of visibility, since the other eigenvector can be derived mathematically. 

In this sense, the results obtained from a characterization method for homogeneous optical elements can be compared with those derived mathematically to test the accuracy of the characterization method. 
Therefore, due to the lack of theoretical foundations with respect of the inhomogeneous matrices, in this study we focus on the optimization of the characterization of homogeneous optical elements with the goal to develop a more robust method that outperforms the only current characterization technique.

Since the case of inhomogeneous Jones matrices is far more complex than that of homogeneous Jones matrices, the method proposed by Garza-Soto et al.~\cite{Garza-Soto} only characterizes homogeneous optical elements~\cite{Gutierrez-Vega2020}. Nevertheless, the method is inefficient, since the search space of the problem is in the realm of the real numbers and so applying a general sampling method is time-consuming. Briefly explained, the method consists of a general search strategy, where the entire search space is evenly sampled to find the eigenvectors of the polarization element. The search is done by rotating two polarization elements to obtain an input polarization state, photographing the interference pattern due to the interaction between the reference and sample beams, and calculating the quality of the interference pattern. This is done for around 90 polarization states. Nonetheless, these 90 measurements were chosen by trial and error with no defined or well-justified criteria. As explained in more detail in Section~\ref{sec:exp_bas}, based on the allowed accuracy of the experimental arrangement, a general sampling method involves 6,000 measurements, making the characterization of any optical element a time-consuming process. Thus, an optimization of the method could reduce the number of measurements and the time required to characterize a homogeneous optical element.

Optimization techniques are used to find a solution to a problem specified by an objective function, where the variables are searched over to find the combination that results in the best objective function value while satisfying the constraints of the problem~\cite{VenterG}. Evolutionary Algorithms (EAs) are optimization techniques based on the theory of evolution, which explains the adaptive changes of species in nature through the survival of the fittest, heredity, and mutation~\cite{Back}. They are all random-based meta-heuristic algorithms that do not require gradient information and typically make use of several points in the search space at a time~\cite{Yu}. Hence, EAs are powerful probabilistic optimization algorithms useful in complex optimization problems~\cite{Beasley2000}. Therefore, using the exploration capabilities of EAs for the characterization of a homogeneous Jones matrix seems to be a promising research area, and it provides an excellent starting point to begin the study of the characterization of Jones matrix of optical elements.

EAs have been previously used in Optics as a new approach to solve different physical problems~\cite{Alander2014}, dating back to the optimization of the design of multi-layer filters~\cite{multilayer} and of silver based heat mirrors~\cite{heatmirrrors1,heatmirrrors2}. More recently, EAs have been applied in polarimetry to optimize the design of a Stokes/Mueller polarimeter. Letnes et al.~\cite{Letnes} optimized the design  of three fast multi-channel Stokes/Mueller polarimeters with close to optimal performance by applying a Genetic Algorithm (GA) based directly on the description of Holland~\cite{holland}. The GA was built on a binary representation with a logic bit negation as a mutation operator, and a multi-point bit crossover as a crossover operator. Similarly, Lo et al.~\cite{Lo} proposed an analytical technique based on the Stokes parameters and the Mueller matrix method to characterize five parameters of anisotropic optical materials by integrating a GA to enable the extraction of the optical properties of a given sample. The GA was based on a real-valued representation with a real-value crossover operator and mutation was performed by inducing a small random perturbation to an individual.

Contrary to the previously presented problems, the problem presented in this study is the first instance of EAs used to optimize the characterization of an optical element using the Jones calculus.
Thus, this study aims to design a new methodology that, taking advantage of EAs, improves the characterization of the Jones matrix of homogeneous optical elements (see Fig.~\ref{fig:homogeneous}) as a first approach to the characterization of polarization elements. In other words, the goal is to reduce the number of measurements and, consequently, the time required in an experimental test. It is expected to find one of the two eigenvectors of a Jones matrix element through an EA given a homogeneous optical element by searching over the fitness search space, so a more efficient sampling must be done in comparison to the general sampling method. In the following, we will show that both EAs implemented, a mutation-based EA and a GA that uses both crossover and mutation operators, out-performed the general sampling method by requiring fewer measurements. Though, the EA performed better by having a perfect rate of convergence on all optical elements characterized here, and still requiring fewer measurements than the GA in our experimental setting.

We now introduce the mathematical background necessary to define the genotype and phenotype of an individual, the fitness function, and the evaluation of an individual's fitness. We continue with the definition of our EA approach, where we present the EA and the GA implemented. Then, we establish the experimental baseline and proceed to showcase the experimentation done using the algorithms. In the end, we discuss the results and conclude the work presented in this study.

\section{Mathematical Background and Physical Phenomena}
In this section, we explain the physical intuition behind the creation of an individual in order to describe its genotype and how it translates into its phenotype. Also, the fitness function is discussed with a general introduction to its representation and how it is measured. Furthermore, in this section, we aim to clarify the complexity of the problem so as to expose the relevance of the EA approach.

Light is an electromagnetic wave that travels through space oscillating transversely to the direction of propagation. The electromagnetic field is composed of a moving electric field $\mathbf{E}(\mathbf{r},t)$ and magnetic field $\mathbf{B}(\mathbf{r},t)$. Mathematically, a monochromatic plane wave can be represented as a vector through the description of its electric field. So, consider a plane wave propagating in a direction described by the unit vector $\mathbf{\hat{k}}$, with angular frequency~$\omega$, velocity~$v$ and an constant initial phase~$\mathbf{\phi_o}$. The electric field vector $\mathbf{E}(\mathbf{r},t)$ of the monochromatic plane wave in space $\mathbf{r}$ and time~$t$ can be described as $\mathbf{E}(\mathbf{r},t) = \text{Re}\left[\mathbf{E_0} e^{i\left(\omega\mathbf{\hat{k}}\cdot \mathbf{r}/v-\omega t- \mathbf{\phi_o} \right)}\right]$,
where~$i$ refers to the imaginary unit defined as $i = \sqrt{-1}$ and $\text{Re}$ refers to the real part of the complex vector.

By convention, the axis of propagation is chosen to be the $z$-axis from negative to positive, so $\mathbf{\hat{k}}=\mathbf{\hat{z}}$. This means that the electric field is restricted to the $x$- and $y$-axis, and so $\mathbf{\hat{k}}\cdot\mathbf{r}=z$ (where $\mathbf{r}$ is the position vector). Additionally, $\mathbf{E}_0$ is a complex vector, so each of its components can be written in polar form. Without loss of generality, we can study the time evolution of the electric field in the plane situated at $z=0$. Thus, $\mathbf{E}(z=0,t)$ can be written as
\begin{equation}
\begin{aligned}
    \mathbf{E}(t) &= \text{Re}
    \begin{bmatrix}
    \mathbf{E}_0
    e^{-i(\omega t + \phi_o)}
    \end{bmatrix} \\
    &= \text{Re}
    \begin{bmatrix}
    \begin{pmatrix}
    A_x e^{-i\phi_x}\\
    A_y e^{-i\phi_y}
    \end{pmatrix}
    e^{-i(\omega t + \phi_o)}
    \end{bmatrix},
    \label{Eq:E(t)}
\end{aligned}
\end{equation}
where $\mathbf{E}_0$ in the first equality is a normalized vector. The time variation of this vector draws what is commonly known as the polarization ellipse or the polarization state of light\footnote{We are using the convention typically used in Optics, where the phase decreases with time and increases with space (i.e., $kz-\omega t$).}. 

\begin{figure}[t!]
    \centering
    \resizebox{0.45\linewidth}{!}{
    \input{figures/tex-images/poincaresphere}
    }
    \caption{Mapping of a small subset of polarization states into the surface of the Poincaré sphere. Notice that the north hemisphere contains right-handed polarization states and the south hemisphere contains left-handed polarization states.}
    \label{fig:poincare}
\end{figure}
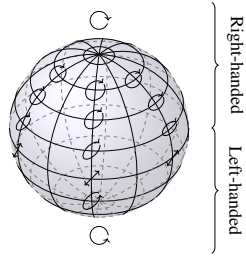

In 1941, Jones~\cite{Jones} developed the Jones vector formalism, where polarization states are represented by a time independent complex vector drawn from Eq.~(\ref{Eq:E(t)}). Meaning that we can describe the polarization state of a polarized beam of light as
\begin{equation}
    \mathbf{E}_0 =
    \begin{pmatrix}
    A_x e^{-i\phi_x}\\
    A_y e^{-i\phi_y}
    \end{pmatrix},
    \label{Eq:JonesVect}
\end{equation}
where the constant initial phase $\phi_o$ has been absorbed by the individual phases of the $x$ and $y$ components. The complex nature of $\mathbf{E}_0$ provides a periodicity, so it allows us to map the variety of polarization states over the surface of a unit sphere known as the Poincaré sphere. The surface of the Poincaré sphere represents all possible polarizations states. Figure~\ref{fig:poincare} shows the Poincaré sphere, where a set of polarization states are shown to illustrate the mapping of different polarizations to the surface of the sphere. We follow the convention where right-handed and left-handed circular polarization states are mapped to the north and south poles, respectively, whereas linear polarization states are located along the equator.

The transformation from the Jones vector to the Poincaré sphere representation is achieved through the Stokes vector parameters $S_1$, $S_2$, and $S_3$. So, with respect to the Jones formalism, the Stokes vector can be written as $S_1 = |E_{0x}|^2-|E_{0y}|^2$,  $S_2 = 2\,\text{Re}\left(E_{0x}^*\, E_{0y}\right)$, and $S_3 = - 2\,\text{Im}\left(E_{0x}^*\, E_{0y}\right)$. In 1956, Pancharatnam~\cite{Pancharatnam} showed that the slow transition from one polarization state to another is accompanied by a phase shift that can be viewed as the geometry of the cycle in the surface of the Poincaré sphere. Therefore, the Stokes parameters offer the link between the Jones mathematical formalism and the geometric nature of the polarization of light. So now that we have defined the description of a polarization state and the relationship between a Jones vector and its position over the Poincaré sphere, we will explain how one polarization state can be transformed to produce another polarization state. 

Polarization states can be manipulated by modifying the components of the Jones vector $\mathbf{E}_0$ in Eq.~(\ref{Eq:JonesVect}), either by changing its amplitude ($A_x$ and $A_y$) or by adding a phase to either component ($e^{-i\phi_x}$ and $e^{-i\phi_y}$). The quarter-wave plate (QWP) and the half-wave plate (HWP) are two well-known polarization elements. The QWP can be used to adjust the eccentricity of the polarization ellipse, and the HWP to adjust the inclination angle of the polarization ellipse. Thus, we can transform the polarization state of a polarized beam of light by letting it traverse through a combination of a QWP and a HWP oriented at different angles, which means that we can explore the entire surface of the Poincaré sphere with the help of these two elements.

Formally, a polarization element is an optical element that can alter the electric field of light to transform its polarization state to another~\cite{Chipman}. Polarization elements also have a mathematical representation in the Jones formalism, in which any polarization element can be described as a $2\times2$ complex-elements matrix. Since we are dealing with polarization elements, the eigenvectors of a Jones matrix, in reality, represent polarization states. Therefore, in the Jones calculus, the eigenvectors are also called eigenpolarizations. So a Jones matrix can be characterized by the identification of the corresponding eigenpolarizations.

On the other hand, polarization elements may be uncharacterized, meaning that the corresponding mathematical description is yet to be defined. As stated previously, homogeneous polarization elements have orthogonal eigenpolarizations, which means that we can derive one eigenpolarization from the other. In other words, we only need to find a single eigenpolarization of the system. As defined by Lopez-Mago et al.~\cite{Lopez-Mago}, we can describe a Jones matrix through its eigenpolarizations as follows. Let $\mathbf{J}$ be a homogeneous Jones matrix and let $\mathbf{q_1} = [q_x; q_y]$ and $\mathbf{q_2} = [-q_y^\ast; q_x^\ast]$ be the corresponding orthonormal eigenvectors of $\mathbf{J}$, where $q_x$, $q_y$ $\in \mathbb{C}$  and ${\rvert q_x \rvert ^2 + \rvert q_y \rvert^2 = 1}$. The eigenvalues of $\mathbf{q_1}$ and $\mathbf{q_2}$ are $\mu_1$ and $\mu_2$, respectively, meaning that $\mathbf{J q_1} = \mu_1\mathbf{q_1}$ and $\mathbf{J q_2} = \mu_2 \mathbf{q_2}$. So, knowing $\mathbf{q_1}$ and $\mathbf{q_2}$ and the corresponding eigenvalues, the Jones matrix \textbf{J} of the polarization element can be written as
\begin{align*}
    \mathbf{J} = \left(\begin{matrix}
    \mu_1\rvert q_x \rvert^2 + \mu_2\rvert q_x \rvert^2 & (\mu_1-\mu_2)q_x q_y^* \\
    (\mu_1-\mu_2)q_x^* q_y & \mu_2\rvert q_x \rvert+\mu_1\rvert q_x \rvert ^2
    \end{matrix}\right).
\end{align*}
Therefore, we need to find an eigenpolarization of a homogeneous optical element to define its mathematical description. In other words, since we know that an eigenvector~$\mathbf{q_{\{1,2\}}}$ of a matrix $\mathbf{J}$ is at most altered by a constant $\mu_{\{1,2\}}$ when the linear transformation $\mathbf{J}\mathbf{q_{\{1,2\}}}$ is applied, we find an eigenvector of a polarization element when a polarization state is altered at most by a scaling factor after the beam of light traverses the element. Fortunately, we can measure the effect of a polarization element over a beam of light by analyzing the polarization state of the beam before and after it interacts with the polarization element. The experimental set-up that allows us to measure this interaction is known as the Mach-Zehnder interferometer, which will be presented in Section~\ref{sub:FitnessFuncDefintion}.

\section{Our Evolutionary Algorithm Approach}
\label{sec:ESApproach}
The previously presented mathematical background is now defined in the context of the EA terminology.

\subsection{Population Representation and Initialization}
\label{sub:PopRepresentation}
To start off the population, we need to define an initial polarization state so that we can produce any polarization state throughout the evolution process. Remembering that we can rotate the QWP and the HWP to transform one polarization state into another, we can set horizontally polarized light~$\mathbf{h}$ as a base state, though we could have chosen any other polarization. The Jones vector $\mathbf{h}$ of the base state and the Jones matrices of the QWP and the HWP are
\begin{gather*}
    \mathbf{h} = 
    \begin{pmatrix}
    1\\0
    \end{pmatrix}, 
    \mathbf{Q}(\alpha) =\frac{1}{\sqrt{2}} \begin{pmatrix}
    1+i\cos(2\alpha) & i\sin(2\alpha) \\
    i\sin(2\alpha) & 1-i\cos(2\alpha)
    \end{pmatrix},\\
    \mathbf{H}(\beta) =\frac{1}{\sqrt{2}} \begin{pmatrix}
    \cos(2\beta) & \sin(2\beta) \\
    \sin(2\beta) & -\cos(2\beta)
    \end{pmatrix},
\end{gather*}
where $\alpha$ and $\beta$ are angles with respect to the fast axes of the wave-plates. The base state $\mathbf{h}$ is transformed as it traverses through the $\mathbf{Q}(\alpha) \mathbf{H}(\beta)$ polarizing stage. So, we can describe any polarization state by rotating the $\mathbf{Q}(\alpha)$ and the $\mathbf{H}(\beta)$, i.e., $\mathbf{v^h}\left(\alpha,\beta\right)=\mathbf{H}(\beta)\, \mathbf{Q}(\alpha)\, \mathbf{h}.$
By setting angles $\alpha$ and $\beta$ to some real constants, the resulting polarization can be described by the Jones vector $\mathbf{v^h}(\alpha,\beta)$ mapped to the Poincaré sphere through the Stokes vector
\begin{align}
    \mathbf{S^h}\left(\alpha,\beta\right) = \begin{bmatrix}
    \cos(2\alpha) \cos(4\beta - 2\alpha) \\
    \cos(2\alpha) \sin(4\beta - 2\alpha)\\
    - \sin(2\alpha)\end{bmatrix}.
    \label{eq:SH}
\end{align}
This equation explicitly shows how angles $\alpha$ and $\beta$ control the resulting polarization state $\mathbf{S^h}$~\cite{Lopez-Mago}. To span all the surface of the Poincaré sphere, both $\alpha$ and $\beta$ must be in the range $[-\pi/4,\pi/4]$. Therefore, angles $\alpha$ and $\beta$ represent an individual's genotype and $\mathbf{S^h}(\alpha,\beta)$ represents an individual's phenotype.

\subsection{Fitness Function Definition}
\label{sub:FitnessFuncDefintion}

\begin{figure}[htp]
\centering
\includegraphics[width=\linewidth]{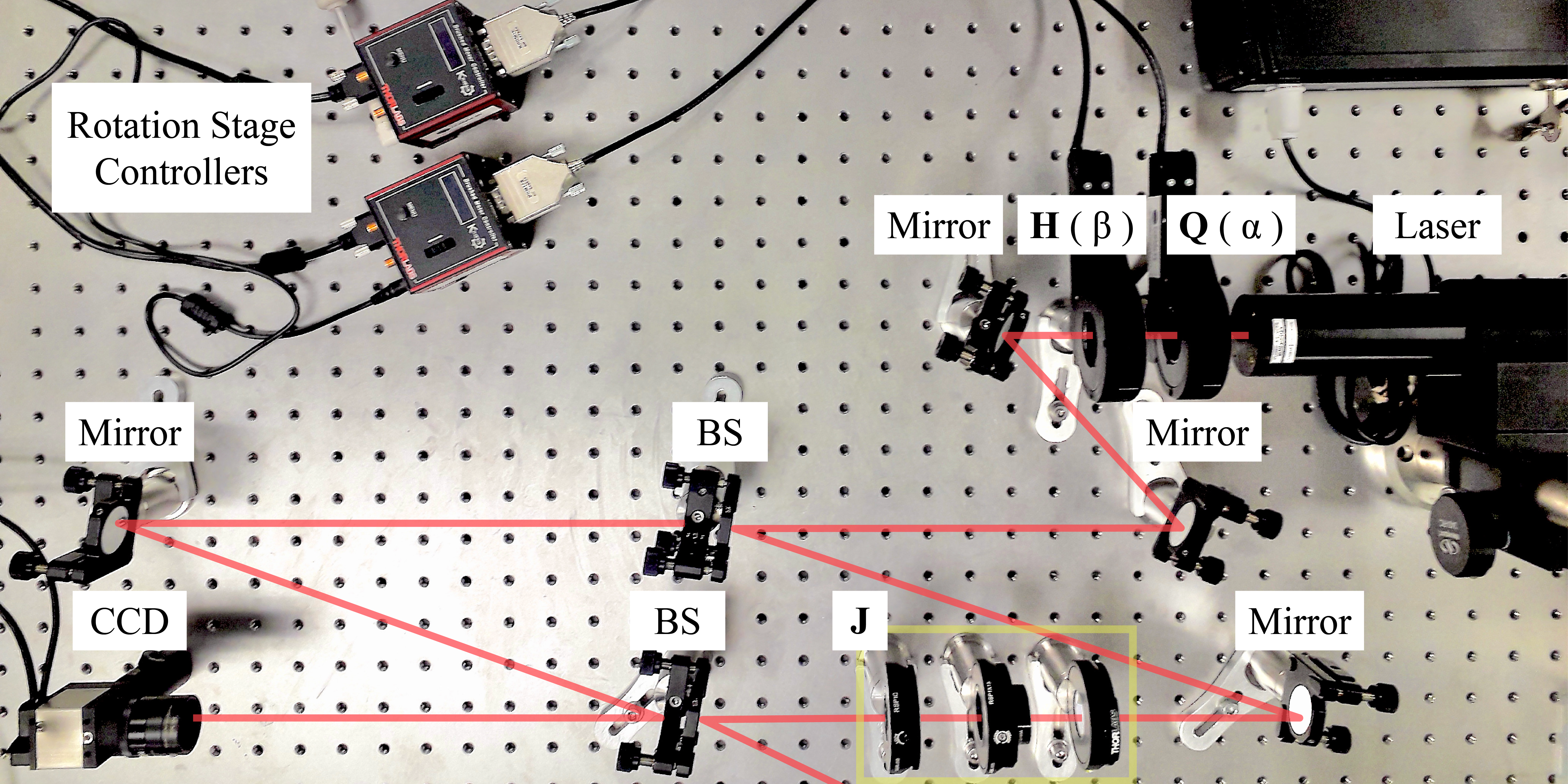}
\caption{The Mach-Zehnder optical arrangement is used to evaluate the fitness of an individual.}
\label{fig:arrangement}
\end{figure}

We can measure the interaction between an input beam (Thorlabs HNL050LB - HeNe) with polarization $\mathbf{S^h_{\boldsymbol{i}}}$ and a homogeneous optical element $\mathbf{J}$ with the help of an optical arrangement known as the Mach-Zehnder interferometer shown in Fig.~\ref{fig:arrangement}~\cite{Garza-Soto}. The $\mathbf{Q}$ (Thorlabs WPQ10M-633 - Ø1") and the $\mathbf{H}$ (Thorlabs WPH10M-633 - Ø1") creates the desired input polarization through the parameters $\alpha$ and $\beta$ using a motorized precision rotation stage controllers (Thorlabs KPRM1E - Ø1"). The beam splitters (BSs) (Thorlabs BSW10 - Ø1") divide the beam with a 50:50 split ratio. The element $\mathbf{J}$ represents the uncharacterized optical element, and the CCD camera (Thorlabs  DCU223M) is used to photograph the interference pattern. The photograph is used to obtain the fitness value as follows. The analitycal experimentation takes into account the limitations of the rotation stage controllers and the definition of the camera. The rotation stage controllers have minimum step of $0.03^{\circ}$ and a accuracy of 0.1\%. Thus, the genotype of an individual is rounded to the nearest interval and a 0.1\% random error taken from the normal distribution is added. Similarly, the CCD camera has a resolution of $1024 \times 768$ pixels, which was also considered.

\begin{figure}[htp]
    \centering
    \subfloat[]{
        \centering
        \includegraphics[width=0.3\linewidth]{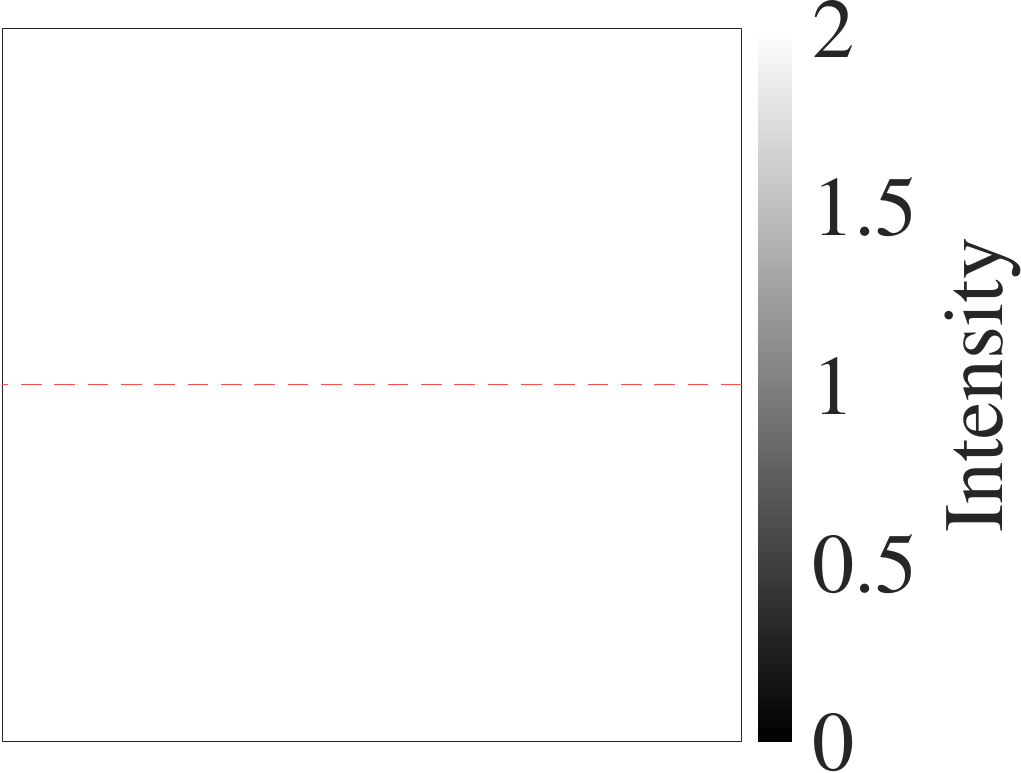}
    \label{fig:IntVis0}
    }
    ~
    \subfloat[]{
        \centering
        \includegraphics[width=0.3\linewidth]{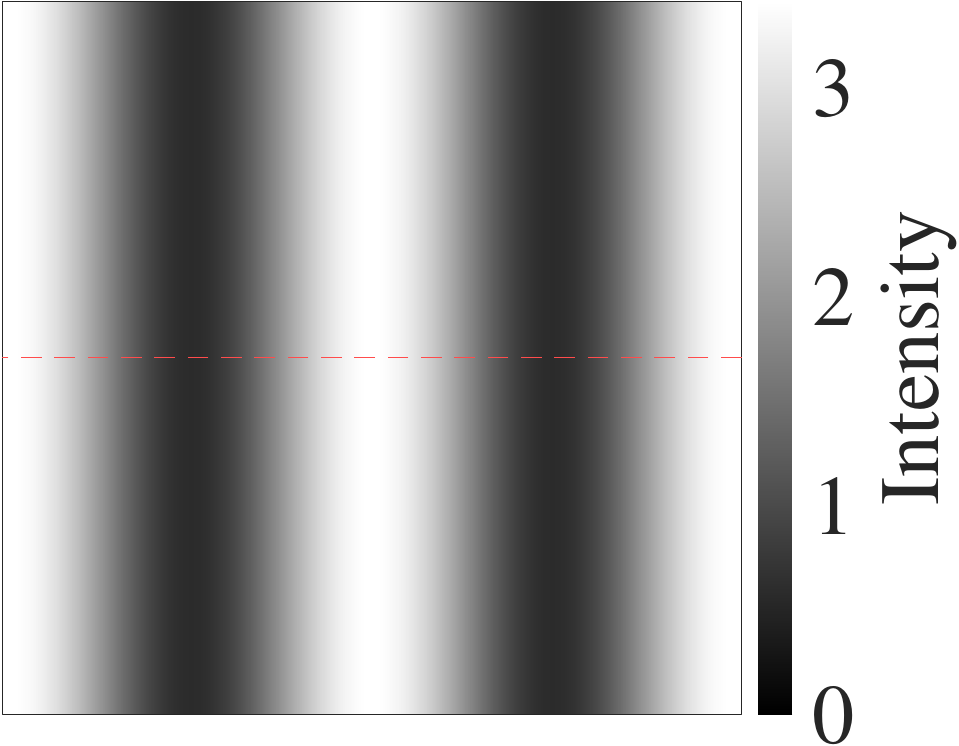}
    \label{fig:IntVis012}
    }
    ~
    \subfloat[]{
        \centering
        \includegraphics[width=0.3\linewidth]{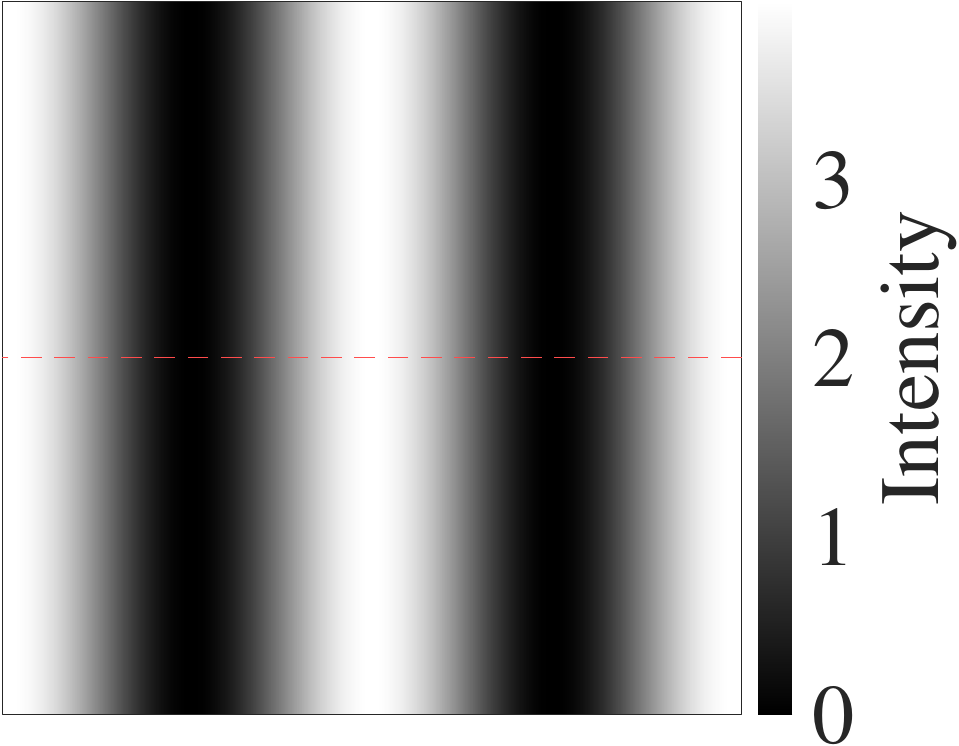}
    \label{fig:IntVis1}
    }
    \caption{Interference patterns observed at the end of the Mach-Zehnder optical arrangement for the interaction between (a)~two beams with orthogonal polarization states ($\visibility = 0$), (b)~two beams with  non-orthogonal polarization states ($0 < \visibility < 1$), and (c)~two beams with the same polarization states ($\visibility=1$). The red dashed line denotes the perpendicular line where the search for the maximum ($P_{\text{max}}$) and minimum ($P_{\text{min}}$) values of the interference pattern is done to obtain the interferometric visibility (see Eq.~(\ref{eq:fitness})).}
    \label{fig:IntVis}
\end{figure}

Recall that the fitness of an individual is determined by the interaction between the polarization states before and after traversing through $\mathbf{J}$, so let $\mathbf{S^{h}_{\boldsymbol{o}}}$ be the state of polarization of the beam after traversing $\mathbf{J}$. We can measure the contrast of the interference between $\mathbf{S^h_{\boldsymbol{i}}}$ and $\mathbf{S^{h}_{\boldsymbol{o}}}$ to evaluate the change of the original polarization state. The contrast in the interference pattern is known as the interferometric visibility, and it can be described as 
\begin{equation}
    \visibility = \frac{P_{\text{max}}-P_{\text{min}}}{P_{\text{max}}+P_{\text{min}}},
    \label{eq:fitness}
\end{equation}
where $P$ is the intensity distribution. The values of $P_{\text{max}}$ and $P_{\text{min}}$ are obtained by searching for the maximum and minimum values in a line perpendicular to the lines of the interference pattern captured by the CCD camera at the end of the experimental arrangement. Fig.~\ref{fig:IntVis} shows a more visual explanation for the obtainment of the value corresponding to the \visibility, where the \visibility is 1 when the two states interacting have the same polarization and 0 when they have orthogonal polarizations. As a side note, the minimum \visibility of an element can be greater than 0 but the maximum \visibility is always 1. Therefore, we aim to find a polarization state with a \visibility of 1, since this means that the polarization of the beam remained the same after traversing through $\mathbf{J}$, i.e., $\mathbf{S^h_{\boldsymbol{i}}} \simeq \mathbf{S^{h}_{\boldsymbol{o}}}$. Hence, the optical arrangement in Fig.~\ref{fig:arrangement} represents the fitness function, and the value of the interferometric visibility represents the fitness of an individual.

\subsection{Evolutionary Algorithm}
The EA now presented is based on the Polynomial Mutation (PM) operator for real-parameter optimization problems in which a polynomial probability distribution is used to perturb the values of a solution~\cite{Deb}. Algorithm~\ref{alg:EAPM} shows the pseudo-code of the implementation. So, just as a reminder, the genotype of an individual is composed of $(\alpha, \beta)$, where $\alpha,\beta \in [-\pi/4, \pi/4]$, the phenotype of an individual is then described by Eq.~(\ref{eq:SH}) and the fitness of an individual can be evaluated using the Mach-Zehnder optical arrangement (Fig.~\ref{fig:arrangement}) and calculating the interferometric visibility defined by Eq.~(\ref{eq:fitness}). Therefore, Algorithm~\ref{alg:EAPM} starts off the initial population $P_{0}$ by generating a set of $\mu$ individuals using pseudo-random values drawn from the standard uniform distribution and afterwards the individuals are evaluated. If an individual in $P_{0}$ is evaluated with the maximum fitness value, the EA finishes its execution. If not, the algorithm continues. So, with the number of offspring per generation set to $\lambda$, $\lambda$ individuals are randomly selected from the~$\mu$ individuals in the population using a discrete uniform distribution in order to create the offspring population $Q_t$. Then, all individuals in $Q_t$ are mutated using the PM operator and then evaluated. The PM operator has a mutation probability of $p_m=1$, meaning that every individual in $Q_t$ is subjected to mutation, and, as explained next, we set a user-defined parameter, denoted as $\eta_m$, to a constant value. At the end of each iteration, the next generation $P_{t+1}$ is created by selecting the most fitted $\mu$ individuals from $P_t\cup Q_t$. If an individual in $P_{t+1}$ has reached the maximum \visibility fitness value or the algorithm has reached the maximum number of fitness evaluations of 6,000, the evolution process terminates. If not, the evolution process continues.

\begin{algorithm}[!ht]
\SetAlgoLined
\KwData{Objective function (\visibility), number of generation ($t$), number of polarization states per generation ($\mu$), number of offspring polarization states per generation ($\lambda$), mutation probability ($p_m$), user-defined parameter $(\eta_m)$.}
\KwResult{Most fitted individual found in the evolution process.}
Initialize population $P_0$ with $\mu$ randomly distributed individuals\;
Evaluate population $P_0$ using \visibility\;
Initialize $t$ to $0$\;
\While{stopping criterion \textbf{not} met}{
    Generate offspring population $Q_t$ of size $\lambda$ from population $P_t$\; 
    Mutate offspring population $Q_t$ using the PM operator with probability $p_m$ and user-defined parameter $\eta_m$\;
    Evaluate offspring using \visibility\;
    Create next generation $P_{t+1}$ of best-fitted $\mu$ individuals from $P_{t}\cup Q_t$\;
    Set $t$ to $t + 1$\;
}
\Return individual with best fitness obtained\;
\caption{Evolutionary Algorithm for Eigenstate Approximation}
\label{alg:EAPM}
\end{algorithm}

Mutation operates independently over one member of the population. Deb and Agrawal~\cite{Deb} suggested a polynomial mutation for real-parameter optimization problems, where a polynomial probability distribution is used to perturb the values of a solution within its vicinity. 
Let $\alpha_i$ be a gene of a randomly selected solution $i$, where $\alpha_i \in [-\pi/4,\pi/4]$. A mutated solution $\alpha'_i$ is created as follows. First, the perturbation $\delta_m$ is calculated as
\begin{align*}
    \delta_m =
    \begin{cases}
    (2u)^{1/(1+\eta_m)}-1, &\text{for} \quad u \leq 0.5,\\
    1- (2(1-u))^{1/(1+\eta_m)} &\text{for} \quad u > 0.5,
    \end{cases}
\end{align*}
where $u$ is a random number in the range $[0,1]$.  Depending on the value of $u$, the mutated gene is calculated as follows
\begin{equation*}
    \alpha'_i = 
    \begin{cases}
    \alpha_i+\delta_m(\alpha_i-\alpha_L) &\text{for} \quad u \leq 0.5,\\
    \alpha_i+\delta_m(\alpha_U-\alpha_i) &\text{for} \quad u > 0.5,
    \end{cases}
\end{equation*}
where $\eta_m$ is the user-defined index parameter and $\alpha_U$ and $\alpha_L$ are the upper and lower bounds of $\alpha$. The same process applies accordingly to $\beta$, in which case $\beta_U$ and $\beta_L$ would correspond to the upper and lower bounds respectively.  Deb and Agrawal~\cite{Deb} concluded that $\eta_m$ induces an effect of perturbation of $\Bigo{(\alpha_U-\alpha_L)/\eta_m}$ in a variable. It is relevant to point out that the probability of mutating outside the upper and lower bounds is zero.

\subsection{Genetic Algorithm}
A second approach was made using a GA with a Simulated Binary Crossover (SBX) operator and the PM operator~\cite{DebSBX}. Algorithm~\ref{alg:GA1P2} presents the structure of the GA. Now, instead of using just mutation as the main variation procedure, in Algorithm~\ref{alg:GA1P2} a pair of individuals is selected randomly using a discrete uniform distribution from the $\mu$ individuals in $P_t$ to undergo crossover and mutation. The probability of recombination is $p_c = 0.5$ and the probability of mutation is $p_m = 1 - p_c = 0.5$. Thus, a random number $u$ is generated and if $u < p_c$, the SBX operator creates two offspring from two randomly selected individuals in $P_t$ and recombines their $\alpha$ and $\beta$ values. Otherwise, $u \leq p_c$ and so the offspring are identical copies of their parents. After, the offspring are subjected to the PM operator with probability $p_m$. The same process continues until $\lambda$ offspring have been produced. Then, Algorithm~\ref{alg:GA1P2} proceeds in the same way as Algorithm~\ref{alg:EAPM}. 

\begin{algorithm}[!ht]
\SetAlgoLined
\KwData{Objective function (\visibility), number of generation ($t$), number of individuals per generation ($\mu$), number of offspring per generation ($\lambda$), crossover probability ($p_c$), mutation probability ($p_m$).}
\KwResult{Most fitted individual found in the evolution process.}
Initialize population $P_0$ with $\mu$ randomly distributed individuals\;
Evaluate population $P_0$ using \visibility\;
Initialize $t$ to $0$\;
\While{stopping criterion \textbf{not} met}{
    Select a multi-set of parents from population $P_t$\;
    Generate offspring population $Q_t$ of size $\lambda$ by using the SBX operator with a probability~$p_c$\;
    Apply the PM operator to the offspring population $Q_t$ with probability $p_m$\;
    Evaluate offspring using \visibility\;
    Create next generation $P_{t+1}$ of best-fitted $\mu$ individuals from $P_{t}\cup Q_t$\;
    Set $t$ to $t + 1$\;
}
\Return individual with best fitness obtained\;
\caption{Genetic Algorithm for Eigenstate Approximation}
\label{alg:GA1P2}
\end{algorithm}

The SBX operator uses a probability distribution created around two parents to produce two offspring resembling the parents. The probability distribution is centered around the parents so that the pair of offspring are more likely to be closer to the parents and the span of the offspring is proportional to the span of the parents. So let $\alpha_i$ and $\alpha_j$ be the pair of genes of two randomly selected individuals $i$ and $j$, where $\alpha_{\{i,j\}}=[-\pi/4,\pi/4]$. Then the offspring's genotypes $\alpha'_{i}$ and $\alpha'_{j}$ are created as follows~\cite{DebSBX2}. First, the perturbation $\delta_c$ is calculated as
\begin{equation*}
    \delta_c = 
    \begin{cases}
    (2u)^{1/(\eta_c+1)} &\text{for}\quad u \leq 0.5,\\
    \left(\frac{1}{2(1-u)}\right)^{1/(\eta_c+1)} &\text{for} \quad u > 0.5,
    \end{cases}
\end{equation*}
where $u$ is a random number in the range $[0,1]$. After, the offspring's genes are calculated as follow
\begin{align*}
    \alpha'_{i} = 0.5((1+\delta_c)\,\alpha_i+(1-\delta_c)\,\alpha_j),
    \\
    \alpha'_{j} = 0.5((1-\delta_c)\,\alpha_i+(1+\delta_c)\,\alpha_j).
\end{align*}
Similarly, the parameters $\beta_i$ and $\beta_j$ of the parents are used to produce the respective genes of the offspring.

\section{Experimentation}

\begin{figure}[htp]
    \centering
    \subfloat[$\mathbf{J_1}\equiv \mathbf{Q}(0) \mathbf{H}(\pi/8)\mathbf{Q}(0)$]{
        \centering
        \includegraphics[width = 0.48\linewidth,trim={35, 5, 35, 35},clip]{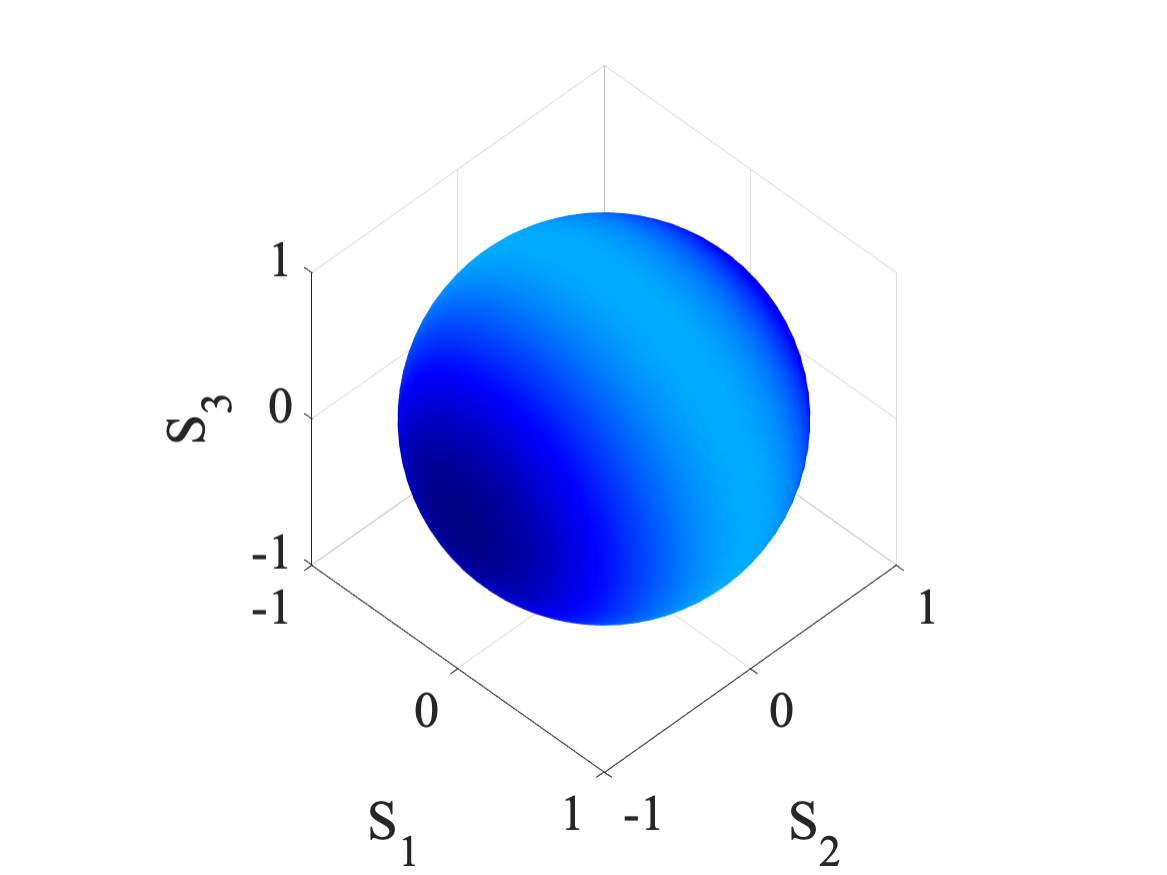}
    \label{fig:FitSpaceQ0Hpi8Q0}
    }
    \subfloat[$\mathbf{J_2} \equiv \mathbf{Q}(0)\mathbf{H} (\pi/4)\mathbf{Q}(0)$]{
        \centering
        \includegraphics[width = 0.48\linewidth,trim={35, 5, 35, 35},clip]{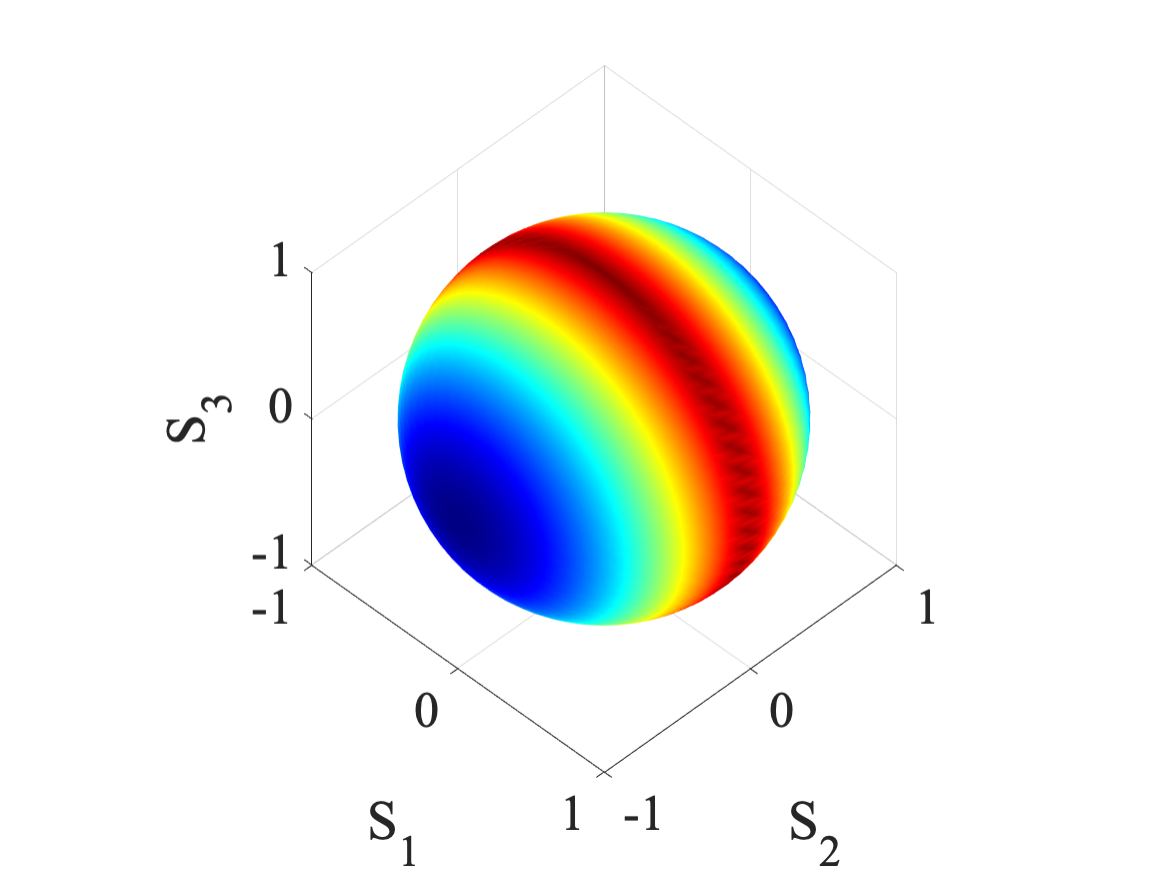}
    \label{fig:FitSpaceQ0Hpi4Q0}
    }
    
    \subfloat[$\mathbf{J_3} \equiv \mathbf{Q}(0)\mathbf{H} (\pi/8)\mathbf{Q}(\pi/8)$]{
        \centering
        \includegraphics[width = 0.48\linewidth,trim={35, 5, 35, 35},clip]{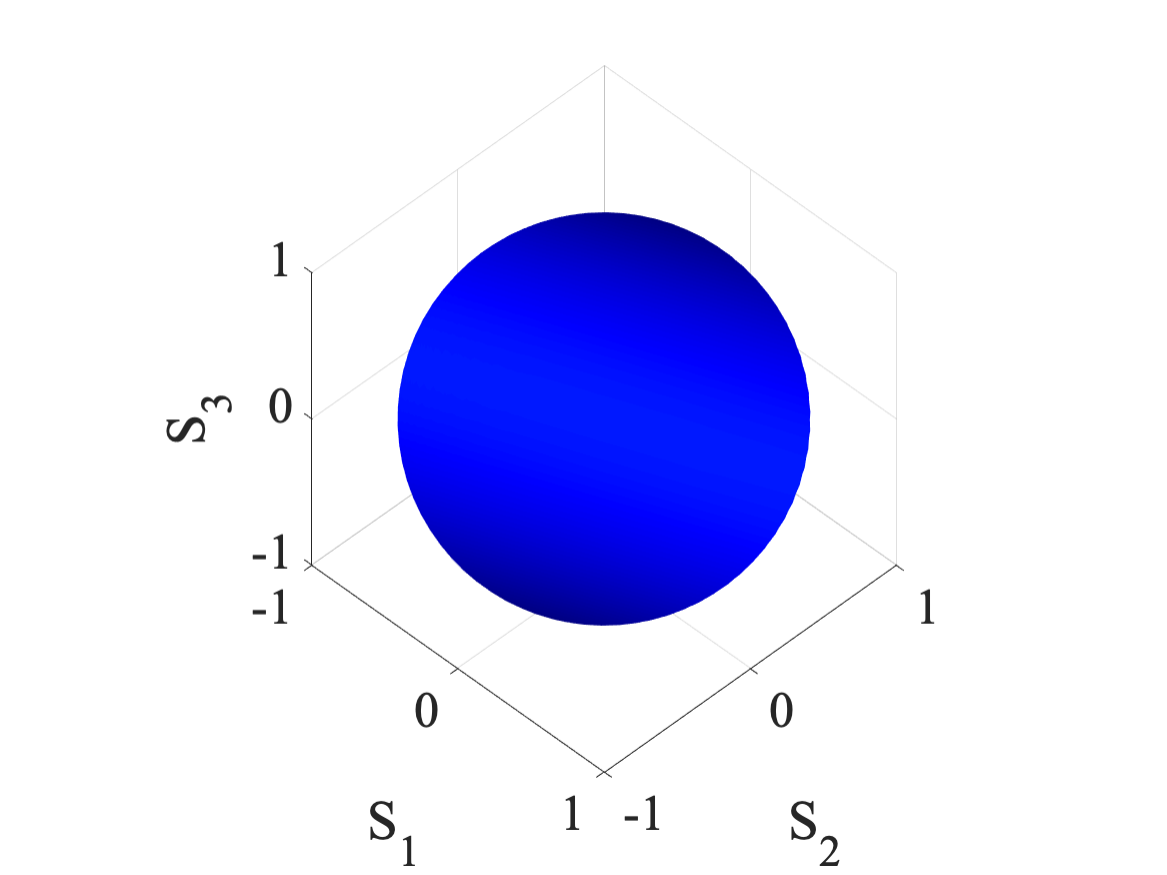}
    \label{fig:FitSpaceQ0Hpi8Qpi8}
    }
    \subfloat[$\mathbf{J_4} \equiv \mathbf{Q}(0)\mathbf{H} (\pi/4)\mathbf{Q}(\pi/4)$]{
        \centering
        \includegraphics[width = 0.48\linewidth,trim={35, 5, 35, 35},clip]{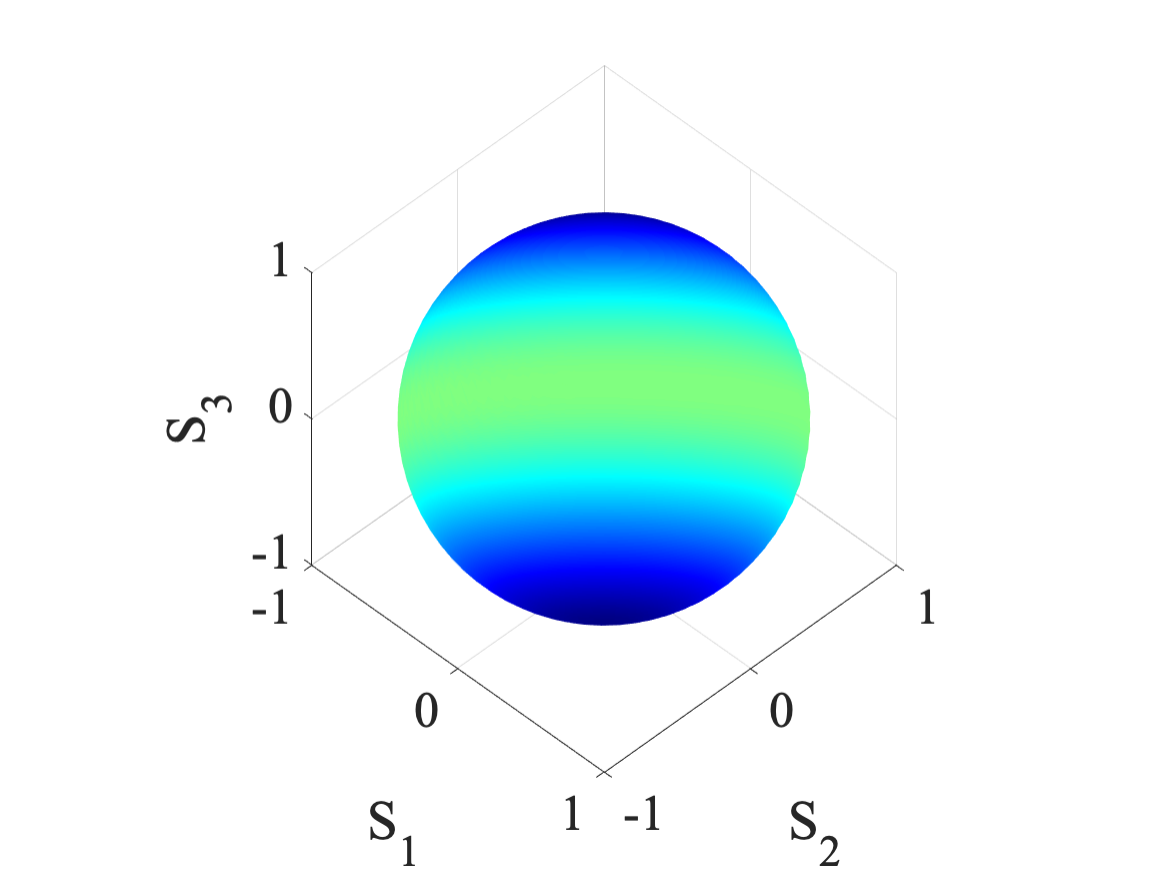}
    \label{fig:FitSpaceQ0Hpi4Qpi4}
    }
    
    \subfloat[$\mathbf{J_5} \equiv \mathbf{Q}(\pi/8)\mathbf{H} (\pi/8)\mathbf{Q}(0)$]{
        \centering
        \includegraphics[width = 0.48\linewidth,trim={35, 5, 35, 35},clip]{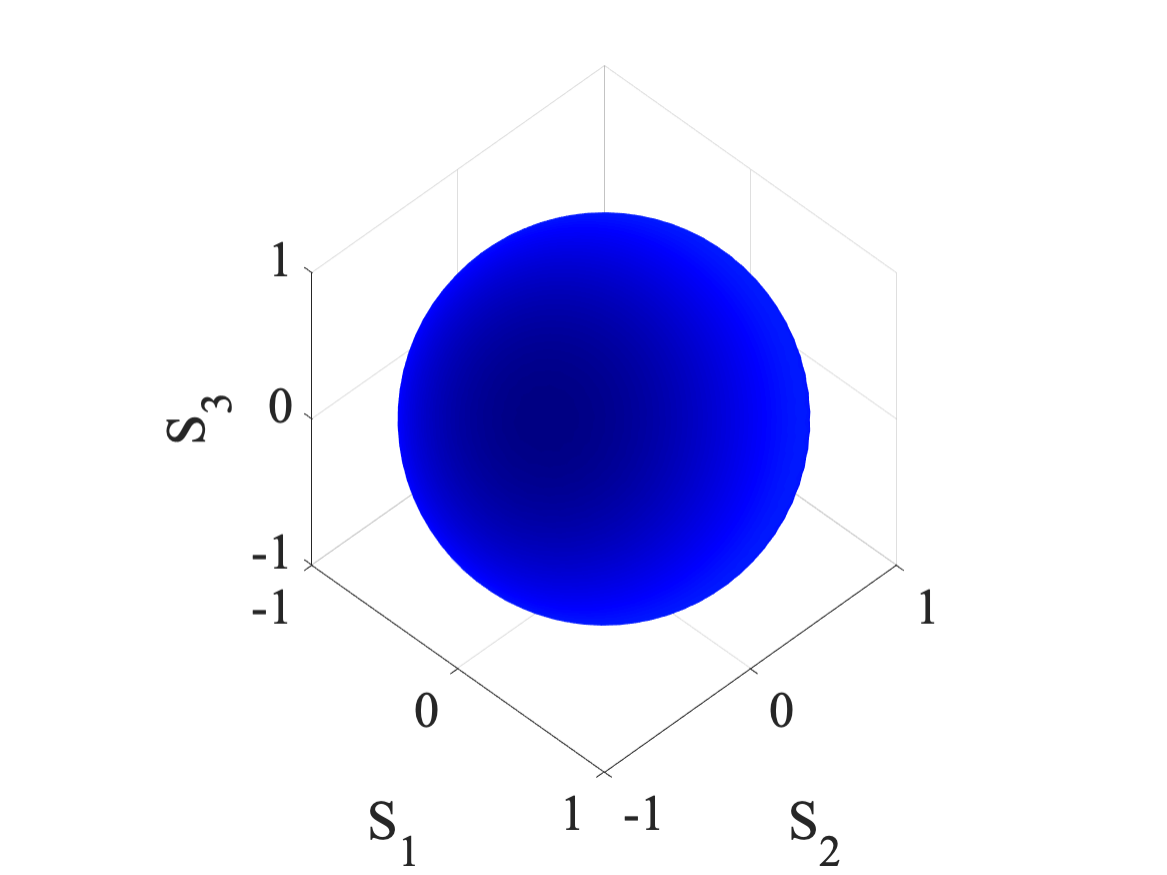}
    \label{fig:FitSpaceQpi8Hpi8Q}
    }
    \subfloat[$\mathbf{J_6} \equiv \mathbf{Q}(\pi/4)\mathbf{H} (\pi/4)\mathbf{Q}(0)$]{
        \centering
        \includegraphics[width = 0.48\linewidth,trim={35, 5, 35, 35},clip]{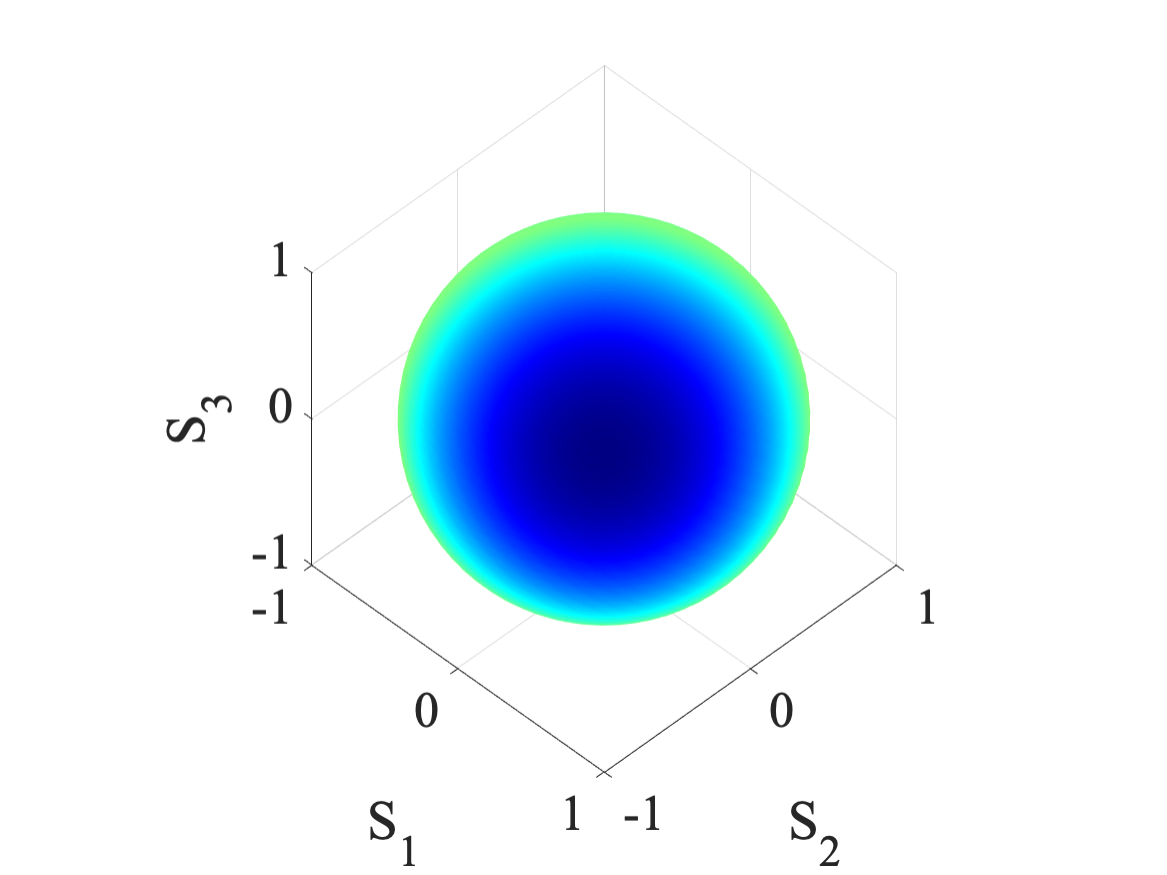}
    \label{fig:FitSpaceQpi4Hpi4Q0}
    }
    
    \subfloat{
    \centering
    \includegraphics[width = .9\linewidth,trim={0, 0, 0, 335},clip]{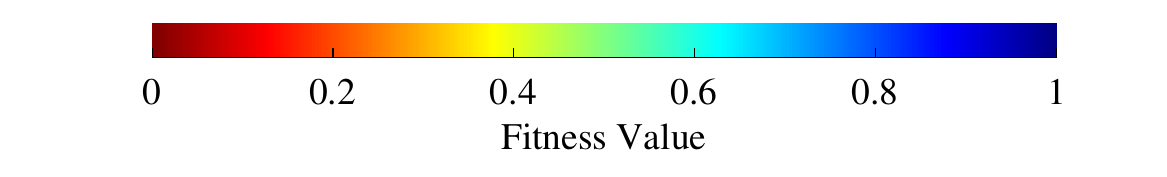}
    }
    \captionsetup{singlelinecheck=off}
    \caption[Fitness search space]{Fitness search space of the optical systems.
    }
    \label{fig:FitnessSpace}
\end{figure}

In order to evaluate the performance of the EA and GA, we make use of known elements to validate the results. So, given that we already have presented the $\textbf{H}$ and the~$\textbf{Q}$ polarization elements, we create a combination of this pair of Jones matrices to produce different optical systems. In this work, we characterize six variations of the optical system $\mathbf{Q}(\theta_1) \mathbf{H}(\theta_2) \mathbf{Q}(\theta_3)$, since by varying the angles we can produce different Jones matrices. The same models were used as in the experimental arrangement (Thorlabs WPQ10M-633 - Ø1" and Thorlabs WPH10M-633 - Ø1"), composed of two multi-order crystalline quartz wave plates to obtain an optical path difference of $\lambda/4$ for $\mathbf{Q}$ and of $\lambda/2$ for $\mathbf{H}$, where here $\lambda$ refers to the wavelength of a beam of light. The search space of the fitness function $\visibility=\Phi_g(\alpha,\beta) \to \mathbb{R}\in[0,1]$ for each system is presented in Fig.~\ref{fig:FitnessSpace}. The elements were selected in order to evaluate the behavior of the algorithms under different optical elements. The elements $\mathbf{J_1}$ and $\mathbf{J_2}$ differ in the gradient and the minimum fitness value, elements $\mathbf{J_4}$ and $\mathbf{J_5}$ differ slightly in the position of the eigenvectors of the systems and in the minimum fitness value, as it is also the case for the elements $\mathbf{J_5}$ and $\mathbf{J_6}$.

First, we will present a brief explanation of the characterization of a system using the traditional sampling method and the required number of evaluations through a general sampling approach, so that later on, we can use this baseline of evaluations to analyze the performance of the EAs with respect to the number of evaluations. Therefore, we now define the experimental baseline.

\subsection{Experimental Baseline}
\label{sec:exp_bas}
\begin{figure}[htp]
    \centering
    \subfloat[]{
        \centering
        \includegraphics[width = 0.45\linewidth]{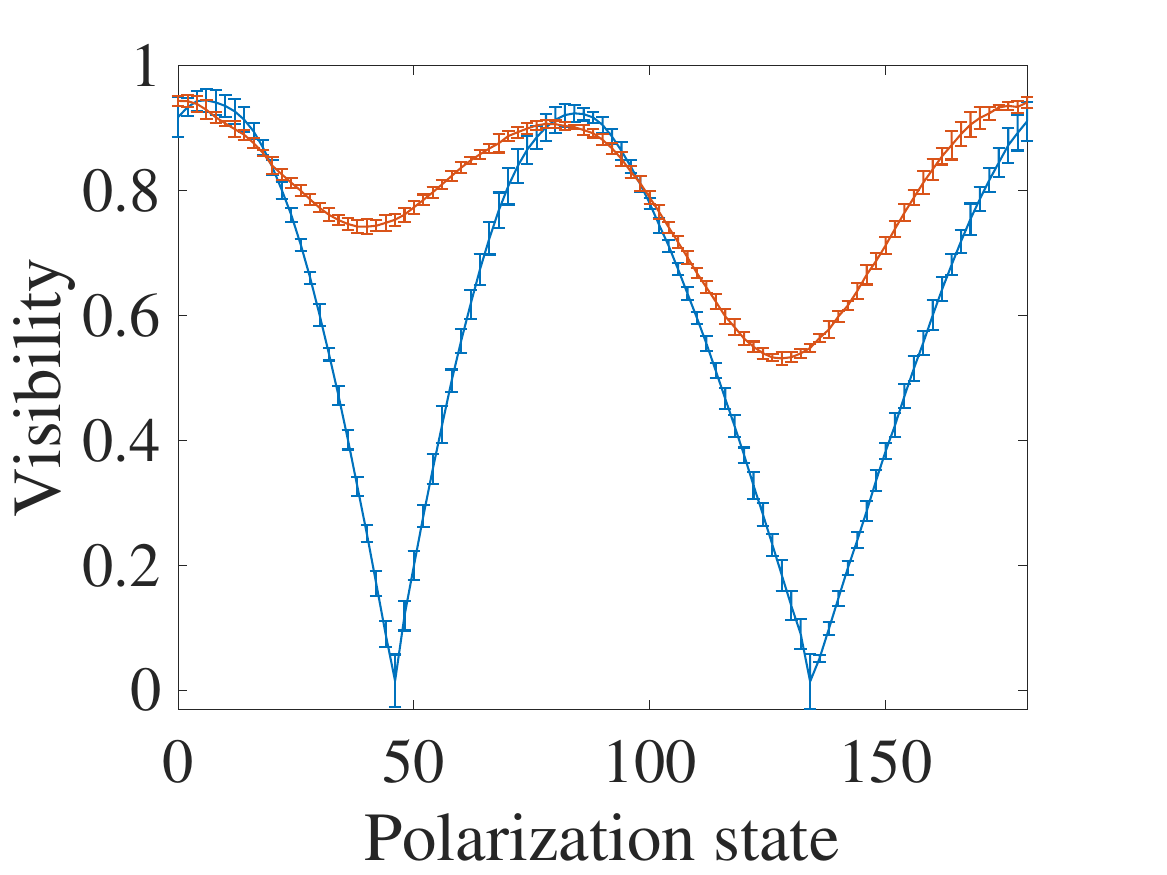}
    \label{fig:ExperimentalJ1J2}
    }
    ~
    \subfloat[]{
        \centering
        \includegraphics[width = 0.48\linewidth]{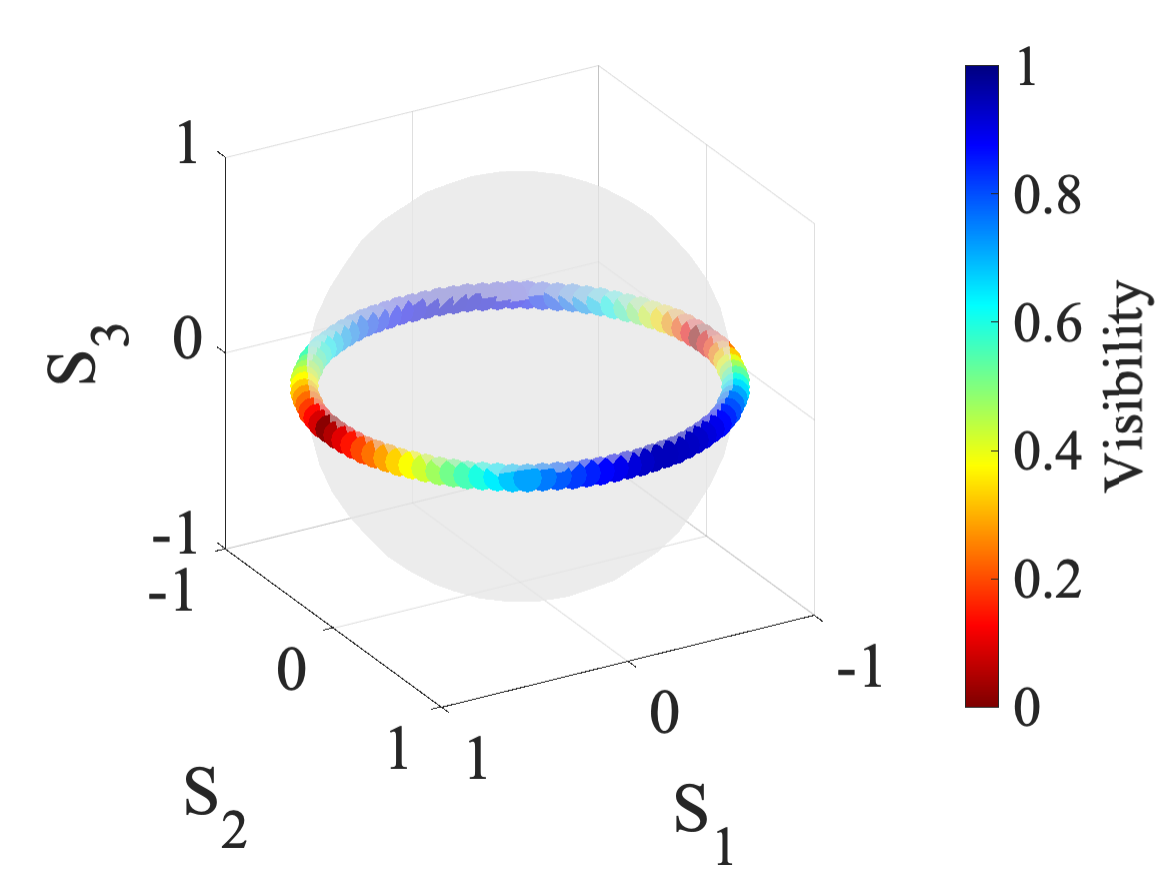}
    \label{fig:ExperimentalEquator}
    }
    \caption{Actual experimental measurements for a traditional search evaluation using (a) an even sampling over the equator with a step of 5 degrees for different systems. The blue line corresponds to the search for the eigenpolarizations of the $\mathbf{J_2}$ system, and the red line corresponds to the search for the eigenpolarizations of the $\mathbf{J_1}$ system. As visual aid, (b) shows the sampling over the Poincaré sphere for the $\mathbf{J_2}$ system.}
    \label{fig:Experimental}
\end{figure}

The traditional search over the Poincaré sphere entails the measurement of a distributed set of polarization states, as seen in Fig.~\ref{fig:Experimental}. In~\cite{Garza-Soto}, the maximum number of feasible measurements is given by the minimum step of the motorized rotation stage of $0.03^{\circ}$ in the range $[0,\pi/2]$ of $\mathbf{H}$. The optimum number of measurements is then the total number of points that the resolution of the engine allows. Nonetheless, that creates a lot of measurements and consequently of experimental time required. To determine a physical optimum of measurements, the authors would have to determine the variance of the phase fluctuations, which is not considered in the theoretical model. Thus, in~\cite{Garza-Soto} a set of 90 measurements evenly distributed over the equator were done respectively to characterize a pair of elements, $\mathbf{J_1}$ and $\mathbf{J_2}$. The input beam was horizontally polarized and the $\mathbf{H}$ was rotated with a step of $2.5^{\circ}$ in the range $[0^{\circ},180^{\circ}]$, given only the $\mathbf{H}$ is necessary to sweep the equator. Each set of measurements was performed ten times to demonstrate the repeatability of the experiment. Therefore, a total of 90 measurements were done to obtain the optimal eigenpolarizations of each system. By trial and error, Garza-Soto et al.~\cite{Garza-Soto} determined that these 90 measurements are sufficient to find the optimum value without compromising the total acquisition time of any given experiment.

The set of measurements can be seen in  Fig.~\ref{fig:ExperimentalJ1J2} or similarly over the Poincare sphere in  Fig.~\ref{fig:ExperimentalEquator} for the element $\mathbf{J_2}$. Notice that the measurements are not ideal due to experimental imperfections (cf.~\cite{Garza-Soto} for details). Again, determining the optimal number of points becomes a more complex problem. So far, it is not well defined how to determine said number of optimal points, since it depends on the sample to be analyzed and the theoretical model that is being used. Determining said optimum for the general sampling method is not part of the scope of this work. Consequently, the number of 90 measurements chosen in~\cite{Garza-Soto} has no theoretical justification and, therefore, the results from the method can't be compared with an EAs based method. Besides, we can see that EAs allow us to avoid the problem of defining a minimum number of measurements. Therefore, we will focus on the experimental constraints imposed by the accuracy of the optical arrangement. Since the motorized rotation stages of the $\mathbf{Q}$ and $\mathbf{H}$ elements have a minimum step of $0.03^{\circ}$, we obtain a total of 6,000 points over the Poincare sphere for a range of $\alpha$ and $\beta$ in$[-\pi/4,\pi/4]$. 

Any other method that seeks to outperform a general search must aim not only to reduce as much as possible the number of measurements, but also to be able to find the optimal eigenpolarizations of each system for any given trial and optical element. In consequence we will make use of the 6,000 measurements to impose a baseline in order to make a first evaluation of both approaches. We now start the analysis on the performance of the EA with the PM operator and then the performance of the GA with the SBX and PM operators.

\subsection{Evolutionary Algorithm Evaluation}
In the following exploratory analysis, we expect to quantify the measurements needed to characterize the Jones matrix of a homogeneous polarizing element through the EA. We will refer to Algorithm~\ref{alg:EAPM} as \mplea, this definition will help us to make direct reference to the EA with the PM operator, selection policy, and specific values for $\mu$ and $\lambda$ when needed. 

Thus, we first examine the behavior of the evolution process by varying the values of $\mu$ and $\lambda$ in $\{2^0,2^1,2^2,2^3\}$ and the value of the parameter $\eta_m$ in $\{20,100\}$, which are the extreme values suggested by Deb and Agrawal~\cite{Deb}. We define our stopping criterion for the EA as of 6,000 measurements, the number of measurements done in the general search to characterize a system that was previously mentioned in Section~\ref{sec:exp_bas}. The reason is simple, we want to observe if the EA is able to obtain an eigenpolarization of each system faster than the general method. Additionally, we accept an error of $10^{-4}$ with respect to the maximum theoretical visibility of~1. The error was defined solely on the basis of the convergence of the analytical simulation. The error introduced by the experimental arrangement will need to be defined in the implementation of the method to account for the random and experimental errors. Thus, the stopping criteria is met when the maximum number of fitness evaluations of 6,000 has been reached, or the maximum \visibility fitness value of $1-10^{-4}$ has been found.

The results of the evaluations for the different combinations of $\mu$ individuals and $\lambda$ offspring with $\eta_m=\{20, 100\}$ are shown in Table~\ref{tab:EA}. On the left side of each sub-table, each cell represents the mean number of evaluations done in the 32 trials for each system and, on the right side, a brief statistical analysis is presented to explore the over-all performance of each EA. Thus, Table~\ref{tab:EA} presents the mean for the subset of evaluations corresponding to each system and the mean, standard deviation (Std) and percentage of outliers for the set of all evaluations of each \mplea and each $\eta_m$. In this context, we define an outlier as a trial that reached the maximum number of evaluations without reaching the threshold fitness value. The over-all best ($\mu+\lambda$) combination for each value of $\eta_m$ is highlighted in each table, where a relatively low number of both mean number of evaluations and percentage of outliers is desired. The standard deviation is included to quantify the effect of the outliers. An important observation is that, in both tables, there is a fair percentage of outliers.

\begin{table*}[htp]
\setlength{\tabcolsep}{4pt}
\renewcommand{\arraystretch}{1.25}
\caption{Results for the \mplea with $\eta_m = \{20,100\}$. On the left side of each table, we present the mean number of evaluations realized to find an eigenvalue of each system $\mathbf{J}_{\boldmath{i}}$, and on the right, we present a statistical analysis for all the measurements corresponding to each possible $\mu+\lambda$ combination.}
\resizebox{\linewidth}{!}{%
\begin{tabular}{@{}cccccccccccc@{}}
\toprule
\multicolumn{10}{c}{{\boldmath$\eta_m = 20$}}\\  \midrule
\multirow{2}{*}{\boldmath$(\mu+\lambda)$} & \multicolumn{6}{c}{\textbf{Mean Number of Evaluations}} & \multicolumn{4}{c}{\textbf{$\mu+\lambda$ Statistical Analysis}}\\
\cmidrule(r){2-7}\cmidrule(l){8-11}
 & $\mathbf{J_1}$ & $\mathbf{J_2}$ & $\mathbf{J_3}$ & $\mathbf{J_4}$ & $\mathbf{J_5}$ & $\mathbf{J_6}$ & \textbf{Mean} & \textbf{Std} & \textbf{Outliers} \\ \midrule
$ 1 + 1 $ & 97.00 & 139.47 & 2102.91 & 1670.28 & 64.53 & 93.44 & 694.60 & 1769.48 & 9.90\% \\
$ 1 + 2 $ & 87.06 & 133.31 & 2112.06 & 1683.38 & 71.56 & 103.56 & 698.49 & 1767.84 & 9.90\% \\
$ 1 + 4 $ & 105.00 & 138.75 & 2117.00 & 1683.75 & 81.13 & 144.75 & 711.73 & 1762.84 & 9.90\% \\
$ 1 + 8 $ & 115.00 & 165.25 & 2127.50 & 1688.00 & 94.25 & 157.50 & 724.58 & 1756.97 & 9.90\% \\
$ 2 + 1 $ & 102.38 & 167.41 & 1176.63 & 1582.03 & 66.41 & 108.81 & 533.94 & 1538.66 & 7.29\% \\
$ 2 + 2 $ & 100.50 & 164.81 & 1181.38 & 1580.44 & 69.75 & 106.38 & 533.88 & 1538.60 & 7.29\% \\
$ 2 + 4 $ & 105.88 & 201.50 & 1188.50 & 1415.50 & 73.88 & 132.25 & 519.58 & 1482.75 & 6.77\% \\
$ 2 + 8 $ & 126.50 & 183.25 & 1032.50 & 1244.00 & 84.25 & 134.50 & 467.50 & 1368.21 & 5.73\% \\
\rowcolor{gray!10!white}[4pt] $ 4 + 1 $ & 102.75 & 151.47 & 637.28 & 877.69 & 81.56 & 117.34 & 328.02 & 1108.32 & 3.65\% \\
$ 4 + 2 $ & 106.94 & 163.94 & 643.06 & 883.88 & 83.88 & 126.75 & 334.74 & 1107.52 & 3.65\% \\
$ 4 + 4 $ & 111.75 & 153.00 & 835.25 & 1231.88 & 100.13 & 130.75 & 427.13 & 1311.27 & 5.21\% \\
$ 4 + 8 $ & 128.00 & 174.00 & 677.00 & 1262.25 & 100.25 & 142.00 & 413.92 & 1243.02 & 4.69\% \\
$ 8 + 1 $ & 126.88 & 201.06 & 672.59 & 1253.34 & 91.66 & 151.25 & 416.13 & 1244.36 & 4.69\% \\
$ 8 + 2 $ & 125.06 & 165.50 & 675.31 & 1267.56 & 100.06 & 141.00 & 412.42 & 1244.35 & 4.69\% \\
$ 8 + 4 $ & 134.13 & 191.00 & 497.38 & 1088.00 & 110.75 & 159.13 & 363.40 & 1102.28 & 3.65\% \\
$ 8 + 8 $ & 163.25 & 221.00 & 522.00 & 1093.75 & 113.00 & 160.50 & 378.92 & 1099.74 & 3.65\% \\
\bottomrule
\end{tabular}
~
\begin{tabular}{@{}cccccccccccc@{}}
\toprule
\multicolumn{10}{c}{{\boldmath$\eta_m = 100$}}\\  \midrule
\multirow{2}{*}{\boldmath$(\mu+\lambda)$} & \multicolumn{6}{c}{\textbf{Mean Number of Evaluations}} & \multicolumn{4}{c}{\textbf{$\mu+\lambda$ Statistical Analysis}}\\
\cmidrule(r){2-7}\cmidrule(l){8-11}
 & $\mathbf{J_1}$ & $\mathbf{J_2}$ & $\mathbf{J_3}$ & $\mathbf{J_4}$ & $\mathbf{J_5}$ & $\mathbf{J_6}$ & \textbf{Mean} & \textbf{Std} & \textbf{Outliers} \\ \midrule
$ 1 + 1 $ & 692.80 & 699.60 & 2545.60 & 2247.00 & 157.40 & 765.50 & 1184.70 & 2209.20 & 17.20\% \\
$ 1 + 2 $ & 707.20 & 716.30 & 2560.10 & 2266.60 & 179.90 & 895.50 & 1220.90 & 2226.60 & 17.70\% \\
$ 1 + 4 $ & 739.90 & 746.10 & 2588.40 & 2308.30 & 212.70 & 929.60 & 1254.20 & 2211.70 & 17.70\% \\
$ 1 + 8 $ & 811.00 & 816.50 & 2644.20 & 2380.30 & 297.80 & 981.50 & 1321.90 & 2182.90 & 17.70\% \\
$ 2 + 1 $ & 508.00 & 514.80 & 1459.80 & 1860.10 & 147.60 & 331.70 & 803.70 & 1829.40 & 10.90\% \\
$ 2 + 2 $ & 523.40 & 534.60 & 1475.90 & 1881.60 & 164.70 & 351.70 & 822.00 & 1824.00 & 10.90\% \\
$ 2 + 4 $ & 556.40 & 562.30 & 1501.00 & 1901.10 & 196.90 & 386.00 & 850.60 & 1814.30 & 10.90\% \\
$ 2 + 8 $ & 615.50 & 625.20 & 1565.30 & 1952.30 & 255.50 & 433.30 & 907.80 & 1797.10 & 10.90\% \\
\rowcolor{gray!10!white}[4pt] $ 4 + 1 $ & 162.70 & 171.70 & 1314.80 & 1347.70 & 163.90 & 354.00 & 585.80 & 1471.00 & 6.80\% \\
$ 4 + 2 $ & 358.60 & 369.60 & 1325.00 & 1197.80 & 174.70 & 255.20 & 613.50 & 1475.20 & 6.80\% \\
$ 4 + 4 $ & 196.40 & 205.00 & 1350.20 & 1558.20 & 198.40 & 384.60 & 648.80 & 1515.10 & 7.30\% \\
$ 4 + 8 $ & 428.50 & 436.50 & 1393.50 & 1438.00 & 264.50 & 420.80 & 730.30 & 1552.30 & 7.80\% \\
$ 8 + 1 $ & 178.41 & 191.66 & 1539.38 & 1604.34 & 183.38 & 203.78 & 650.16 & 1514.38 & 7.29\% \\
$ 8 + 2 $ & 197.50 & 214.50 & 1526.44 & 1432.63 & 191.44 & 213.25 & 629.29 & 1461.83 & 6.77\% \\
$ 8 + 4 $ & 213.50 & 225.13 & 1537.63 & 1649.25 & 200.75 & 230.75 & 676.17 & 1510.02 & 7.29\% \\
$ 8 + 8 $ & 245.25 & 264.50 & 1568.50 & 1655.50 & 239.75 & 270.75 & 707.38 & 1502.29 & 7.29\% \\
\bottomrule
\end{tabular}
}
\label{tab:EA}
\end{table*}

\subsection{Genetic Algorithm Evaluation}
In this section we will refer to Algorithm~\ref{alg:GA1P2} as \mplga to make direct reference to the GA with the SBX and PM operators, selection policy, and specific values for $\mu$ and $\lambda$ when needed. We first examine the behavior of the evolution process by varying the values of $\mu$ and $\lambda$ in $\{2^1,2^2,2^3\}$ and of the parameters $\eta_m$ and $\eta_c$ in $\{20,100\}$. Again, the stopping criteria for the algorithm is finding an individual with a \visibility fitness value of $1-10^{-4}$ or reaching the maximum number of fitness evaluations of 6,000. 

The results of the evaluations for the different combinations of $\mu$ individuals and $\lambda$ offspring with ${\eta_m=\{20, 100\}}$ and ${\eta_c=\{20, 100\}}$ are shown in Table~\ref{tab:GA}. In the left side of each sub-table, each cell represents the mean number of evaluations done in the 32 trials for each system and, on the right side, a brief statistical analysis is presented to explore the over-all performance of each \mplga. Thus, Table~\ref{tab:GA} presents the mean number of evaluations for the subset of evaluations corresponding to each system and the mean, standard deviation and percentage of outliers for the set of all evaluations for each \mplga, $\eta_c$ and $\eta_m$.

\begin{table*}[htp]
\setlength{\tabcolsep}{4pt}
\renewcommand{\arraystretch}{1.25}
\caption{Results for the \mplga with ${\eta_c = \{20,100\}}$ and ${\eta_m = \{20,100\}}$. On the left side of each table, we present the mean number of evaluations realized to find an eigenvalue of each system $\mathbf{J}_{\boldmath{i}}$, and on the right, we present a statistical analysis for all the measurements corresponding to each possible $\mu+\lambda$ combination.}
\centering
\resizebox{\linewidth}{!}{%
\begin{tabular}{@{}cccccccccccc@{}}
\toprule
\multicolumn{10}{c}{{\boldmath$\eta_c = 20$}, {\boldmath$\eta_m = 20$}}\\  \midrule
\multirow{2}{*}{\boldmath$(\mu+\lambda)$} & \multicolumn{6}{c}{\textbf{Mean Number of Evaluations}} & \multicolumn{4}{c}{\textbf{$\mu+\lambda$ Statistical Analysis}}\\
\cmidrule(r){2-7}\cmidrule(l){8-11}
 & $\mathbf{J_1}$ & $\mathbf{J_2}$ & $\mathbf{J_3}$ & $\mathbf{J_4}$ & $\mathbf{J_5}$ & $\mathbf{J_6}$ & \textbf{Mean} & \textbf{Std} & \textbf{Outliers} \\ \midrule
$ 2 + 2 $ & 235.69 & 297.19 & 1224.00 & 1689.25 & 132.13 & 179.81 & 626.34 & 1523.73 & 7.29\% \\
$ 2 + 4 $ & 291.25 & 390.88 & 1236.38 & 1665.25 & 124.25 & 229.25 & 656.21 & 1534.44 & 7.29\% \\
$ 2 + 8 $ & 274.75 & 396.75 & 1242.75 & 1481.00 & 146.50 & 227.50 & 628.21 & 1466.66 & 6.77\% \\
$ 4 + 2 $ & 158.75 & 243.50 & 1235.94 & 1467.31 & 119.81 & 223.38 & 574.78 & 1471.18 & 6.77\% \\
$ 4 + 4 $ & 201.88 & 237.00 & 1240.50 & 1666.25 & 141.38 & 244.50 & 621.92 & 1516.97 & 7.29\% \\
$ 4 + 8 $ & 223.00 & 322.00 & 1072.00 & 1666.25 & 134.75 & 227.00 & 607.50 & 1464.76 & 6.77\% \\
$ 8 + 2 $ & 191.31 & 246.75 & 1076.88 & 1111.63 & 132.81 & 237.06 & 499.41 & 1297.60 & 5.21\% \\
\rowcolor{gray!10!white}[4pt] $ 8 + 4 $ & 195.75 & 343.63 & 714.38 & 621.75 & 169.88 & 221.13 & 377.75 & 942.69 & 2.60\% \\
$ 8 + 8 $ & 220.25 & 355.50 & 1268.00 & 841.00 & 165.25 & 208.50 & 509.75 & 1228.02 & 4.69\% \\
\bottomrule
\end{tabular}
~
\begin{tabular}{@{}cccccccccccc@{}}
\toprule
\multicolumn{10}{c}{{\boldmath$\eta_c = 100$}, {\boldmath$\eta_m = 20$}}\\  \midrule
\multirow{2}{*}{\boldmath$(\mu+\lambda)$} & \multicolumn{6}{c}{\textbf{Mean Number of Evaluations}} & \multicolumn{4}{c}{\textbf{$\mu+\lambda$ Statistical Analysis}}\\
\cmidrule(r){2-7}\cmidrule(l){8-11}
 & $\mathbf{J_1}$ & $\mathbf{J_2}$ & $\mathbf{J_3}$ & $\mathbf{J_4}$ & $\mathbf{J_5}$ & $\mathbf{J_6}$ & \textbf{Mean} & \textbf{Std} & \textbf{Outliers} \\ \midrule
$ 2 + 2 $ & 218.44 & 323.00 & 1225.44 & 1848.94 & 119.13 & 227.25 & 660.36 & 1570.92 & 7.81\% \\
$ 2 + 4 $ & 333.63 & 431.00 & 1235.50 & 1668.75 & 129.63 & 196.13 & 665.77 & 1531.49 & 7.29\% \\
$ 2 + 8 $ & 261.50 & 401.00 & 1258.25 & 1479.75 & 141.25 & 240.00 & 630.29 & 1468.21 & 6.77\% \\
$ 4 + 2 $ & 195.06 & 267.50 & 1254.50 & 1481.81 & 109.75 & 261.69 & 595.05 & 1467.15 & 6.77\% \\
$ 4 + 4 $ & 196.00 & 253.63 & 1262.75 & 1497.38 & 133.63 & 176.25 & 586.60 & 1467.22 & 6.77\% \\
$ 4 + 8 $ & 224.00 & 335.50 & 1245.50 & 1708.25 & 152.50 & 259.75 & 654.25 & 1508.15 & 7.29\% \\
$ 8 + 2 $ & 201.56 & 281.44 & 1070.88 & 1330.81 & 127.44 & 206.69 & 536.47 & 1356.65 & 5.73\% \\
\rowcolor{gray!10!white}[4pt] $ 8 + 4 $ & 229.38 & 316.75 & 890.38 & 810.00 & 167.50 & 225.63 & 439.94 & 1093.68 & 3.65\% \\
$ 8 + 8 $ & 214.00 & 336.50 & 1112.25 & 992.25 & 153.50 & 212.25 & 503.46 & 1232.65 & 4.69\% \\
\bottomrule
\end{tabular}}
\newline
\vspace*{.2 cm}
\newline
\resizebox{\linewidth}{!}{%
\begin{tabular}{@{}cccccccccccc@{}}
\toprule
\multicolumn{10}{c}{{\boldmath$\eta_c = 20$}, {\boldmath$\eta_m = 100$}}\\  \midrule
\multirow{2}{*}{\boldmath$(\mu+\lambda)$} & \multicolumn{6}{c}{\textbf{Mean Number of Evaluations}} & \multicolumn{4}{c}{\textbf{$\mu+\lambda$ Statistical Analysis}}\\
\cmidrule(r){2-7}\cmidrule(l){8-11}
 & $\mathbf{J_1}$ & $\mathbf{J_2}$ & $\mathbf{J_3}$ & $\mathbf{J_4}$ & $\mathbf{J_5}$ & $\mathbf{J_6}$ & \textbf{Mean} & \textbf{Std} & \textbf{Outliers} \\ \midrule
$ 2 + 2 $ & 626.19 & 639.31 & 1573.81 & 1970.63 & 278.00 & 449.94 & 922.98 & 1795.57 & 10.94\% \\
$ 2 + 4 $ & 654.50 & 663.88 & 1606.25 & 2151.75 & 315.13 & 469.63 & 976.85 & 1823.39 & 11.46\% \\
$ 2 + 8 $ & 713.50 & 727.50 & 1517.00 & 2221.75 & 397.00 & 545.50 & 1020.38 & 1769.46 & 10.94\% \\
$ 4 + 2 $ & 268.60 & 284.20 & 1428.70 & 1500.10 & 263.70 & 444.40 & 698.30 & 1454.20 & 6.80\% \\
$ 4 + 4 $ & 285.00 & 304.50 & 1460.50 & 1588.80 & 312.50 & 475.30 & 737.80 & 1454.10 & 6.80\% \\
$ 4 + 8 $ & 348.80 & 361.80 & 1511.70 & 1711.00 & 542.30 & 545.70 & 836.90 & 1533.00 & 7.80\% \\
$ 8 + 2 $ & 288.20 & 299.70 & 1619.10 & 1285.30 & 260.60 & 281.50 & 672.40 & 1397.10 & 6.30\% \\
\rowcolor{gray!10!white}[4pt] $ 8 + 4 $ & 304.40 & 325.60 & 1456.60 & 1353.80 & 291.10 & 304.50 & 672.70 & 1337.90 & 5.70\% \\
$ 8 + 8 $ & 336.00 & 351.80 & 1505.20 & 1520.50 & 313.50 & 333.20 & 726.70 & 1391.50 & 6.30\% \\
\bottomrule
\end{tabular}
~
\begin{tabular}{@{}cccccccccccc@{}}
\toprule
\multicolumn{10}{c}{{\boldmath$\eta_c = 100$}, {\boldmath$\eta_m = 100$}}\\  \midrule
\multirow{2}{*}{\boldmath$(\mu+\lambda)$} & \multicolumn{6}{c}{\textbf{Mean Number of Evaluations}} & \multicolumn{4}{c}{\textbf{$\mu+\lambda$ Statistical Analysis}}\\
\cmidrule(r){2-7}\cmidrule(l){8-11}
 & $\mathbf{J_1}$ & $\mathbf{J_2}$ & $\mathbf{J_3}$ & $\mathbf{J_4}$ & $\mathbf{J_5}$ & $\mathbf{J_6}$ & \textbf{Mean} & \textbf{Std} & \textbf{Outliers} \\ \midrule
$ 2 + 2 $ & 625.60 & 640.90 & 1576.40 & 1973.20 & 276.50 & 447.60 & 923.40 & 1795.20 & 10.90\% \\
$ 2 + 4 $ & 654.40 & 670.60 & 1609.00 & 2159.10 & 316.70 & 481.00 & 981.80 & 1822.70 & 11.50\% \\
$ 2 + 8 $ & 723.50 & 740.50 & 1681.80 & 2222.20 & 399.00 & 554.50 & 1053.60 & 1802.20 & 11.50\% \\
$ 4 + 2 $ & 273.10 & 292.10 & 1436.80 & 1504.70 & 266.80 & 453.90 & 704.60 & 1452.90 & 6.80\% \\
$ 4 + 4 $ & 294.60 & 310.30 & 1463.00 & 1538.20 & 317.20 & 483.30 & 734.40 & 1450.60 & 6.80\% \\
$ 4 + 8 $ & 355.30 & 371.30 & 1515.50 & 1609.30 & 365.50 & 541.30 & 793.00 & 1442.70 & 6.80\% \\
\rowcolor{gray!10!white}[4pt] $ 8 + 2 $ & 279.56 & 295.69 & 1471.31 & 1315.81 & 265.25 & 281.00 & 651.44 & 1341.68 & 5.73\% \\
$ 8 + 4 $ & 302.25 & 318.13 & 1636.75 & 1317.00 & 304.38 & 314.00 & 698.75 & 1391.90 & 6.25\% \\
$ 8 + 8 $ & 346.75 & 363.75 & 1657.75 & 1380.25 & 328.75 & 364.50 & 740.29 & 1388.96 & 6.25\% \\
\bottomrule
\end{tabular}}
\label{tab:GA}
\end{table*}

\subsection{Results and Discussion}
The results from the evaluation of both algorithms are promising. It is evident that the \mplea requires fewer evaluations to find an eigenpolarization than the  \mplga, possibly because the search space is not complex enough to require the diversity provided by the GA. Furthermore, lower $\eta_m$ and $\eta_c$ values seem to be more beneficial to the search since less evaluations were required to find an eigenvector in both the \mplea and the GA. Nonetheless, the number of outliers stand out in both the \mplea and the \mplga, meaning that the search is somehow being limited. A possibility is that the analytical boundaries being imposed may be stagnating the search by creating a non-continuous search space. Therefore, when an eigenpolarization falls in or near the boundaries of the search space, the evolution process is falling pray of our limited definition of the search space. Thus, it would also be of benefit the exploration of an alternate definition of our problem to provide a search space which reflects the continuity of the fitness function without prejudicing its simplicity. Overall, the number of evaluations required by both algorithms were less than in the general search. Plus, the \mplea stands out by requiring less evaluations to find an eigenpolarization. The best case obtained was of $328.02$ mean number of evaluations with the \xpxea{4}{1} and $\eta_m=20$.

\subsection{Extended PM and SBX bounds}
Given that it is relevant that both EA and GA methods be able to characterize a system in any given trial and for any optical element, we explore a redefinition of the PM and SBX operators to reduce the number of outliers. Therefore, we explore the mimicry of continuity in our search space by allowing the $\alpha$ and $\beta$ values of an individual's genotype to mutate beyond the previously imposed boundaries with the aim of preventing the search from getting stuck when the eigenvectors are located in or near the boundaries. Thus, we extend the PM and the SBX boundaries of $\alpha, \beta \in [-\pi/4, \pi/4]$ to $\alpha, \beta \in [-\pi/2,\pi/2]$. In other words, we initialize our individuals randomly over the Poincaré sphere in the same manner as before, but we now extend the search space by loosening the limits in the variation operators in order to prevent the creation of the previously imposed boundaries. We can imagine this as the creation of a second wrapping over the sphere to simulate the periodic nature of the search space. 

Thus, we create the same experimental set-up but redefine the variation operators bounds. This time, the best case for the \mplea was obtained with $\eta_m = 100$ and for the \mplga with $\eta_m = 100$ and $\eta_c = 20$. Due to space restrictions, we only present the tables containing said best results, Table~\ref{tab:EA-extended} shows the results for the \mplea with $\eta_m = 100$ and Table~\ref{tab:GA-extended} shows the results for the \mplga with $\eta_m = 100$ and $\eta_c = 20$.  We can see that the main goal of reducing the number of outliers was achieved, in both cases the percentage of outliers dropped down to 0\%. Plus, the mean number of evaluations were also reduced. Over-all, the \xpxea{2}{1} was the best performing algorithm with $103.16$ average evaluations.

\begin{table}[htp]
\setlength{\tabcolsep}{4pt}
\renewcommand{\arraystretch}{1.25}
\caption{Results for the \mplea with $\eta_m = 100$ and extended PM bounds. On the left, we present the mean number of evaluations realized to find an eigenvalue of each system $\mathbf{J}_{\boldmath{i}}$, and on the right, we present a statistical analysis for all the measurements corresponding to each possible combination of $\mu+\lambda$.}
\resizebox{\linewidth}{!}{%
\begin{tabular}{@{}cccccccccccc@{}}
\toprule
\multirow{2}{*}{\boldmath$(\mu+\lambda)$} & \multicolumn{6}{c}{\textbf{Mean Number of Evaluations}} & \multicolumn{4}{c}{\textbf{$\mu+\lambda$ Statistical Analysis}}\\
\cmidrule(r){2-7}\cmidrule(l){8-11}
 & $\mathbf{J_1}$ & $\mathbf{J_2}$ & $\mathbf{J_3}$ & $\mathbf{J_4}$ & $\mathbf{J_5}$ & $\mathbf{J_6}$ & \textbf{Mean} & \textbf{Std} & \textbf{Outliers} \\ \midrule
$ 1 + 1 $ & 105.91 & 116.91 & 106.81 & 127.34 & 96.59 & 130.66 & 114.04 & 73.80 & 0.00\% \\
$ 1 + 2 $ & 120.56 & 127.63 & 119.75 & 144.50 & 104.00 & 142.38 & 126.47 & 80.53 & 0.00\% \\
$ 1 + 4 $ & 137.50 & 147.50 & 149.00 & 156.75 & 124.13 & 172.75 & 147.94 & 91.86 & 0.00\% \\
$ 1 + 8 $ & 185.00 & 198.25 & 208.00 & 222.75 & 178.75 & 221.25 & 202.33 & 121.51 & 0.00\% \\
\rowcolor{gray!10!white}[4pt] $ 2 + 1 $ & 105.13 & 118.72 & 90.38 & 104.03 & 96.69 & 104.00 & 103.16 & 61.88 & 0.00\% \\
$ 2 + 2 $ & 118.50 & 139.56 & 101.94 & 121.50 & 106.81 & 121.38 & 118.28 & 66.98 & 0.00\% \\
$ 2 + 4 $ & 129.75 & 145.75 & 122.38 & 149.88 & 124.88 & 145.25 & 136.31 & 83.52 & 0.00\% \\
$ 2 + 8 $ & 169.75 & 186.75 & 174.50 & 186.75 & 163.25 & 183.50 & 177.42 & 103.53 & 0.00\% \\
$ 4 + 1 $ & 105.25 & 125.09 & 100.06 & 116.84 & 105.59 & 118.16 & 111.83 & 64.56 & 0.00\% \\
$ 4 + 2 $ & 116.06 & 129.75 & 114.31 & 125.44 & 115.06 & 137.25 & 122.98 & 72.40 & 0.00\% \\
$ 4 + 4 $ & 135.00 & 152.75 & 137.00 & 141.38 & 129.25 & 153.13 & 141.42 & 79.25 & 0.00\% \\
$ 4 + 8 $ & 158.75 & 177.50 & 155.75 & 178.50 & 186.25 & 170.25 & 171.17 & 109.66 & 0.00\% \\
$ 8 + 1 $ & 122.69 & 143.91 & 143.59 & 177.94 & 111.84 & 127.25 & 137.87 & 79.44 & 0.00\% \\
$ 8 + 2 $ & 144.06 & 170.94 & 148.69 & 173.44 & 121.38 & 141.69 & 150.03 & 72.56 & 0.00\% \\
$ 8 + 4 $ & 143.75 & 180.00 & 172.00 & 197.50 & 129.50 & 165.88 & 164.77 & 97.79 & 0.00\% \\
$ 8 + 8 $ & 162.50 & 193.50 & 199.50 & 244.25 & 142.75 & 166.25 & 184.79 & 116.11 & 0.00\% \\
\bottomrule 
\end{tabular}}
\label{tab:EA-extended}
\end{table}

\begin{table}[htp]
\setlength{\tabcolsep}{4pt}
\renewcommand{\arraystretch}{1.25}
\caption{Results for the \mplga with $\eta_m = 100$, $\eta_c = 20$ and extended PM bounds. On the left, we present the mean number of evaluations realized to find an eigenvalue of each system $\mathbf{J}_{\boldmath{i}}$, and on the right, we present a statistical analysis for all the measurements corresponding to each possible combination of $\mu+\lambda$.}
\resizebox{\linewidth}{!}{%
\begin{tabular}{@{}cccccccccccc@{}}
\toprule
\multirow{2}{*}{\boldmath$(\mu+\lambda)$} & \multicolumn{6}{c}{\textbf{Mean Number of Evaluations}} & \multicolumn{4}{c}{\textbf{$\mu+\lambda$ Statistical Analysis}}\\
\cmidrule(r){2-7}\cmidrule(l){8-11}
 & $\mathbf{J_1}$ & $\mathbf{J_2}$ & $\mathbf{J_3}$ & $\mathbf{J_4}$ & $\mathbf{J_5}$ & $\mathbf{J_6}$ & \textbf{Mean} & \textbf{Std} & \textbf{Outliers} \\ \midrule
$ 2 + 2 $ & 183.06 & 210.50 & 182.50 & 205.69 & 186.63 & 233.06 & 200.24 & 129.02 & 0.00\% \\
$ 2 + 4 $ & 209.13 & 235.88 & 204.88 & 234.88 & 210.13 & 246.13 & 223.50 & 139.73 & 0.00\% \\
$ 2 + 8 $ & 257.00 & 275.75 & 236.25 & 276.50 & 257.50 & 282.00 & 264.17 & 176.17 & 0.00\% \\
\rowcolor{gray!10!white}[4pt] $ 4 + 2 $ & 175.69 & 197.50 & 188.81 & 198.56 & 179.56 & 185.94 & 187.68 & 121.49 & 0.00\% \\
$ 4 + 4 $ & 180.13 & 225.50 & 200.75 & 209.88 & 199.25 & 209.25 & 204.13 & 133.27 & 0.00\% \\
$ 4 + 8 $ & 217.25 & 254.75 & 243.00 & 272.50 & 240.50 & 255.25 & 247.21 & 151.11 & 0.00\% \\
$ 8 + 2 $ & 185.63 & 211.75 & 218.56 & 215.69 & 169.00 & 205.13 & 200.96 & 103.00 & 0.00\% \\
$ 8 + 4 $ & 190.75 & 222.00 & 230.88 & 226.25 & 172.63 & 220.25 & 210.46 & 109.74 & 0.00\% \\
$ 8 + 8 $ & 225.75 & 259.00 & 253.75 & 268.00 & 204.50 & 229.25 & 240.04 & 123.57 & 0.00\% \\
\bottomrule
\end{tabular}}
\label{tab:GA-extended}
\end{table}

\section{Conclusion}
EAs have had an increasing interest in the past decade both as a research subject and as a method for solving real-world problems. In this paper, we have developed a methodology using EAs to optimize the characterization of the Jones matrices of homogeneous optical elements. We implemented an \mplea with the PM operator and \mplga with the SBX and PM operators to reduce the number of evaluations required by the general search method. The experimental results showed that both algorithms are able to find the maximum visibility and the corresponding values for $\alpha$ and $\beta$ of the search space with fewer evaluations, and with a high rate of convergence, which signifies the development of an efficient and reliable method. Therefore, we have shown that EAs are effective and useful in the optimization of the search for the eigenvectors of homogeneous optical elements. The present analysis has shown that the usage of EAs in the area of polarimetry is a promising research area and as future research, we would like to keep exploring the effect of other parameters like $\eta_c$ and $\eta_m$ in order to decrease even further the number of evaluations, and the general application of EAs on the more complex case of inhomogeneous optical elements, for which no method of characterization currently exists.

\section*{Acknowledgments}
The authors would like to thank the Tecnológico de Monterrey, and the Consejo Nacional de Ciencia y Tecnología - CONACYT (the Mexican National Council for Science and Technology) for the financial support under the CVU 1007204. D.L.M. acknowledges support from CONACYT (Grants No. 299057, No. 295239, and No. APN2016-3140).

\vfill

\bibliographystyle{IEEEtran}
\bibliography{bibliography}

\end{document}

%% file: figures/tex-images/poincaresphere.tex
\usetikzlibrary{decorations.pathreplacing}

\newcommand\pgfmathsinandcos[3]{%
  \pgfmathsetmacro#1{sin(#3)}%
  \pgfmathsetmacro#2{cos(#3)}%
}
\newcommand\LongitudePlane[3][current plane]{%
  \pgfmathsinandcos\sinEl\cosEl{#2} 
  \pgfmathsinandcos\sint\cost{#3} 
  \tikzset{#1/.style={cm={\cost,\sint*\sinEl,0,\cosEl,(0,0)}}}
}
\newcommand\LatitudePlane[3][current plane]{%
  \pgfmathsinandcos\sinEl\cosEl{#2} 
  \pgfmathsinandcos\sint\cost{#3} 
  \pgfmathsetmacro\yshift{\cosEl*\sint}
  \tikzset{#1/.style={cm={\cost,0,0,\cost*\sinEl,(0,\yshift)}}} %
}
\newcommand\DrawLongitudeCircle[2][1]{
  \LongitudePlane{\angEl}{#2}
  \tikzset{current plane/.prefix style={scale=#1}}
  \pgfmathsetmacro\angVis{atan(sin(#2)*cos(\angEl)/sin(\angEl))} %
  \draw[current plane] (\angVis:1) arc (\angVis:\angVis+180:1);
  \draw[current plane,dashed,black!50] (\angVis-180:1) arc (\angVis-180:\angVis:1);
}
\newcommand\DrawLatitudeCircle[2][2]{
  \LatitudePlane{\angEl}{#2}
  \tikzset{current plane/.prefix style={scale=#1}}
  \pgfmathsetmacro\sinVis{sin(#2)/cos(#2)*sin(\angEl)/cos(\angEl)}
  \pgfmathsetmacro\angVis{asin(min(1,max(\sinVis,-1)))}
  \draw[current plane] (\angVis:1) arc (\angVis:-\angVis-180:1);
  \draw[current plane,dashed,black!50] (180-\angVis:1) arc (180-\angVis:\angVis:1);
}


\begin{tikzpicture} 
\def\R{2.5} 
\def\angEl{35} 
\shade[ball color = blue!20!white, opacity=0.3] (0,0) circle (\R);
\foreach \t in {-80,-60,...,80} { \DrawLatitudeCircle[\R]{\t} }
\foreach \t in {-5,-35,...,-175} { \DrawLongitudeCircle[\R]{\t} }

\draw[<-, scale=.3, rotate=45] (3.2,2.7) arc (0:350:1 and .8);
\draw[<-, scale=.3, rotate=45] (1.2,.9) arc (0:350:1 and .7);
\draw[<-, scale=.3,rotate=45] (-1,-1) arc (0:340:1 and .5);
\draw[<->, shift={(-.2,-1.4)}] (-0.2,-0.2) -- (.2,.2);
\draw[->, scale=.3,rotate=45] (-4.3,-4.3) arc (0:340:1 and .5);

\draw[<-, scale=.25, rotate=45] (7.5,1.1) arc (0:350:1 and .8);
\draw[<-, scale=.25, rotate=45] (7.5,-2.5) arc (0:350:1 and .7);
\draw[<-, scale=.25,rotate=45] (6.5,-5.6) arc (0:340:1 and .5);
\draw[<->, shift={(2.05,-.84)}] (-0.15,-.22) -- (.15,.22);

\draw[<-, scale=.25, rotate=45] (2,7.5) arc (0:350:1 and .8);
\draw[<-, scale=.25, rotate=45] (-1.4,7.3) arc (0:350:1 and .7);
\draw[<-, scale=.25, rotate=45] (-4.6,6.35) arc (0:340:1 and .5);
\draw[<->, shift={(-2.28,-.623)}] (-.1,-.2) -- (.1,.2);

\draw[<-, scale=.25] (1,12) arc (0:340:1 and 1);
\draw[->, scale=.25] (1,-12) arc (0:340:1 and 1);

\draw [decorate,decoration={brace,amplitude=5pt,mirror,raise=4pt},yshift=0pt]
(3,0) -- (3,3.5) node [black,midway,xshift=0.8cm] {\rotatebox{-90}{\Large{Right-handed}}};
\draw [decorate,decoration={brace,amplitude=5pt,raise=4pt},yshift=0pt]
(3,0) -- (3,-3.5) node [black,midway,xshift=0.8cm] {\rotatebox{-90}{\Large{Left-handed}}};

\end{tikzpicture}
